\documentclass[10pt,reqno]{amsart}

\usepackage{graphicx}
\usepackage{dcolumn}
\usepackage{bm}
\usepackage{mathrsfs}
\usepackage{amsrefs}
\usepackage{hyperref}
\usepackage{amssymb}
\usepackage{amsaddr}
\usepackage{latexsym}
\usepackage{amsfonts}
\newcommand{\be}{\begin{equation}}
\newcommand{\ee}{\end{equation}}
\newcommand{\ba}{\begin{eqnarray}}
\newcommand{\ea}{\end{eqnarray}}
\newcommand{\nn}{\nonumber}

\newcommand{\xa}{{\bf X}_a } 
\newcommand{\xab}{{\bf X}_{ab} }

\newcommand{\mf}{\mathcal F}

\newcommand{\cal}{\mathcal}

\begin{document}


\title{Elastic bending energy: a variational approach}
\author{Riccardo Capovilla}
\address{
Departamento de F\'{\i}sica, Cinvestav-IPN, Av. Instituto Polit\'ecnico Nacional 2508,
col. San Pedro Zacatenco, 07360, Gustavo A. Madero, Ciudad de M\'exico, M\'exico}


\begin{abstract}
Geometric continuum models for fluid lipid membranes are considered  using classical field theory, 
within a covariant variational approach. The approach is cast as a higher-derivative 
Lagrangian formulation of continuum classical field theory, and it   
can be seen as a covariant version of the field theoretical variational approach that uses the height  representation.
This novel Lagrangian formulation is presented first for a generic reparametrization invariant geometric model,  deriving its equilibrium
condition equation, or shape equation, and its linear and angular stress tensors,  using  the classical  
Canham-Helfrich elastic bending energy for illustration. 
The robustness of the formulation is established by extending it to  the presence of external forces, and to the case of heterogenous 
lipid membranes, breaking reparametrization invariance.  In addition, a useful and compact general expression for the second variation of the free energy is obtained 
within the Lagrangian formulation, as a first step towards the study of the stability of membrane configurations.  
The simple structure of the expressions derived for the basic  entities that appear in the mechanics of a lipid membrane
is a direct consequence of the well established power of a Lagrangian variational approach. 
The paper is self-contained, and  it is meant to provide, besides a new framework, also a convenient  introduction to the mechanics 
of lipid membranes. 
\end{abstract}

\maketitle

\section{Introduction}
Many soft materials are described at mesoscopic scales by geometric models \cite{Kamien-rmp}. Fluid lipid membranes provide a paradigmatic  case, where the geometric object  of interest is a surface, and effective curvature models 
describe remarkably well their physical behaviour  \cite{LN,Handbook,Seifert,Boal,Brown}. 
As the basic building materials in  cellular structures and partitions,  lipid membranes are of central interest in biology and  biophysics \cite{Alberts,Mou,MVF}.
In a solvent, under appropriate conditions, the lipid molecules form
spontaneously  closed membranes  of molecular thickness, a few nanomemeters,
and size much larger of the order of fifty nanometers and up to several micras, so that a 
description of their shapes in terms of a surface is sensible.   At scales larger than molecular, the membrane can be considered as a continuum two-dimensional fluid, and the fluidity of the membrane, or negligible shear, implies that  the description must be reparametrization invariant. Reparametrization invariance is better understood from an active
point of view, rather than from the equivalent  passive point of view of mere change of coordinates: there is no energetic cost in moving a point on the surface representing the membrane. In addition, the lipid molecules
arrange themselves in a bilayer made of two leaflets, to hide their hydrophobic `tails'
from the solvent, and this `bilayer trick' is of  fundamental importance in the physical properties of both artificial lipid membranes and
bio-membranes \cite{Mou}.
The amazingly rich morphology generated by the bilayer architecture of lipid membranes  is captured
by an elastic bending   free energy that is constructed as a sum of geometric scalars. The
resulting curvature geometric models, and in particular the classic Canham-Helfrich free energy  \cite{Can,Hel,Evans}, provide a remarkably  effective description of the configurations and of the mechanical response  of physical lipid membranes \cite{SL1,Seifert,Deserno15}. 

The richness of the morphology and of the shape transformations of lipid membranes is matched by the variety of different approaches to membrane mechanics.
In soft matter physics, one approach to the problem of the determination of the equilibrium
configurations is to use a variational principle,  and a systematic functional variation to  derive the equations  of mechanical equilibrium by minimizing
the geometric free energy, using tools of differential geometry \cite{Hel87,Hel89,Ou,Tu,CGS03}. 
Other approaches to the general description of the static properties of lipid membranes have been put forward,
with different perspectives, for example in elasticity  theory \cite{Jenkins,Steigeq,Deseri}, from the point of view of the membrane nematic microstructure \cite{Napolieq}, 
in applied mathematics using phase field calculus \cite{phase},
in biomechanics \cite{Mercker}, and 
in the differential geometry of  surfaces \cite{Nitsche}.

In applications, the full mathematical covariant approach is not always needed, and  to make analytical
 progress, one specializes to axisymmetric configurations, or surfaces of revolution. 
 In this simplification, the variational principle 
 can be interpreted as an analogue of Hamilton's principle of classical mechanics, with the free energy
 interpreted as an action describing the trajectory of a  fictitious particle, and the parameter along the surface
 contour playing the role of a time parameter \cite{Deuling,ES,Seifert}. The problem is reduced to a finite number of degrees of freedom, but it is still highly non trivial, involving higher derivative non-linear ordinary differential equations
 \cite{Turecent}. 
  In field theory, a different type of simplification is obtained by the use of the height, or Monge, representation,
  where the surface is
represented by a height function, that depends on the two coordinates of a  planar reference configuration \cite{Boal}.
The variational principle becomes a field theoretical model for a scalar field in two dimensions \cite{Jerusalem}. 
 The simplification occurs for  almost planar surfaces, when the surface deviates slightly from the planar reference configuration, assuming no overhangs. The equilibrium equations become linear, and the planar symmetry of the reference
 configurations 
can be exploited to expand the height function in Fourier modes. To examine the short wavelength regime of thermal fluctuations, for example, this is adequate, and the geometric non-linearities can be added as interactions \cite{Seifert,Safram,Peliti}.

 In this paper, our  purpose is  to show that  the differential geometric variational approach to the mechanics of lipid membranes can be organized as a covariant field theoretical Lagrangian formulation, offering a new perspective.
This formulation generalizes the
Lagrangian formulation for axisymmetric configurations to field theory, and it removes the restriction of a 
reference planar configuration in
 the variational approach  that uses the height representation.  The Lagrangian formulation is of the higher-derivative, sometimes called generalized, type, because of
 the dependance of the energy on curvature, and therefore on second derivatives of the field variables. 
The field variables are the shape functions that define the surface that represents the membrane,
and the Lagrangian phase space is given by their covariant derivatives.  The role of the Lagrangian function is played
by the membrane energy density, and its integral over the surface, the free energy, is the corresponding action
functional.
As a mechanical system,
the free energy can be considered as a classical field theory, although in fact it is closer to a 
relativistic field theory, when using the symmetry of reparametrization invariance as guidance.
The variational principle produces naturally the covariant Euler-Lagrange equations that define equilibrium, or shape
equation as it has come to be known for geometric models. 
The simple structure of these equations does not appear to be known in the literature, and it is a 
direct consequence of the Lagrangian formulation.
The covariance of the variational approach, without 
any early gauge fixing, and the symmetry of the free energy under rigid motions in space, using Noether's theorem, give readily general expression in terms of the phase space partial derivatives of the energy density for
 the membrane  linear and angular stresses. Also in this case, the simple expressions obtained are an
added benefit of the Lagrangian formulation.  
 In a classical mechanical analogy, the linear stress tensor can be considered as  the analogue of the linear momentum of an acceleration dependent  point-like particle, and it plays an equally
 fundamental role.  
 The linear and the angular stress tensors provide a remarkably efficient
tool in the evaluation of forces and torques within the membrane, and are especially valuable when 
considering interactions with  substrates or with objects embedded in the membrane. This is the subject of the
extensive review by Deserno \cite{Deserno15}, to which we will refer the reader repeatedly in this paper.
Our strategy is to first focus on reparametrization invariant geometric models, that describe homogenous membranes,
in order to introduce the Lagrangian formulation.
For illustration, we specialize to a lipid membrane described by the Canham-Helfrich  free energy, that we use
as a concrete example.   For this geometric model, we offer an alternative derivation of 
results  obtained previously in the differential geometric variational
approach, and organize them within the Lagrangian formulation. 
Next, we show how in the covariant  Lagrangian formulation one can include external forces, and
the possibility of describing in a natural way inhomogeneities in the membrane lipid composition. Also in in these
cases, the classical mechanics analogy  is useful.
The Lagrangian formulation is also put to use in the derivation of a general expression for the second variation
of the free energy, where it is  shown that it can be expressed almost solely  in terms of the Hessians of the
energy density. This is a first necessary step in the study of the stability and the perturbations of equilibrium
configurations, and the second variation is expressed in a form suitable for applications, also in the
presence of interactions.

The paper is organized as follows. In order to make the paper self-contained, Sect. 2 contains a brief summary of the basic notions of the 
geometry of surfaces, with emphasis on the mathematics needed in the
classical field formalism, and in particular on the shape functions and its derivatives. 
In Sect. 3, the elastic energy density for lipid membranes is introduced, as 
an expansion in the derivatives of the shape functions. 
The covariant  Lagrangian formulation is the subject of Sect. 4,
where the fist variation of the free energy  is used to derive the equilibrium shape equation, 
the linear and the angular stress tensors for lipid membranes. 
The Canham-Helfrich  model is used both to illustrate the formalism and for the sake of comparison with previous treatments.
In sect. 5, the formulations is extended to consider heterogenous lipid membranes, breaking reparametrization
invariance, and the presence of external forces.
In sect 6, the second variation of the
energy density is derived within the covariant Lagrangian approach, as a first step in the 
study of the issue of the stability and the shape transformations of lipid membranes.
 We conclude in Sect. 7 with some final remarks.
 According to Lagrangian tradition, the reader will not find any figures in this text.

\section{Geometry}

In order to make the paper self-contained, and to establish our notation, in this section we
describe briefly the geometry of a two-dimensional surface. For an introduction to the local geometry of surfaces, tailored to 
problems in soft matter physics and lipid membranes see \cite{Kamien-rmp,Deserno15,powers};
for a mathematical monograph see {\it e.g.} do Carmo \cite{doCarmo}.

A lipid membrane is modeled as a two-dimensional orientable surface $\mathbb{S}$ embedded in three-dimensional
Euclidean space. This surface $\mathbb{S}$ is specified locally,
in parametric form, by three shape functions
${\bf X} = ( X^1,X^2,X^3)$, as
\[
{\bf x} = {\bf X} (\xi^a ),
\]
where the coordinates ${\bf x} = x^\mu = (x^1 , x^2 , x^3 )$
describe a point in space ($\mu,\nu, \dots, = 1,2,3$),
$\xi^a = (\xi^1 , \xi^2 )$ are arbitrary coordinates on the surface ($a,b, \dots = 1,2$).

The shape functions are the classical fields of interest, and our main purpose in this section is to describe
the geometrical content of their surface derivatives. The first surface derivatives are,
geometrically, the two tangent vectors to the surface 
\be
{\bf X}_a = \partial_a {\bf X} = {\partial {\bf X} (\xi^a)  \over \partial \xi^a}.
\ee
The metric induced on the surface $\mathbb{S}$
by the embedding  is defined by their inner product
\be
g_{ab} = {\bf X}_a \cdot {\bf X}_b = \delta_{\mu\nu} X^\mu_a X^\nu_b.
\ee
Latin indices are lowered and raised with $g_{ab}$ and its inverse
$g^{ab}$, respectively, where the inverse is defined by $g^{ac} g_{cb} =\delta^a_b$.
Greek indices are lowered and raised with the Kronecker delta. Whenever it does
not lead to confusion, the boldface notation is used. 

The induced metric describes the intrinsic geometry of the surface. For the description of the extrinsic geometry,
the unit normal ${\bf n}$ to the surface $\mathbb{S}$ is
defined implicitly, up to a sign,  by
\be
{\bf n} \cdot {\bf X}_a = 0, \quad \quad {\bf n}\cdot {\bf n} = 1.
\ee
It is customary, for a closed surface, to choose the outwards pointing one as positive.
Alternatively, since the ambient space is three-dimensional, the unit normal can be written as
the cross product 
\be
{\bf n} = {{\bf X}_1 \times {\bf X}_2 \over | {\bf X}_1 \times {\bf X}_2| } = {1 \over 2 \sqrt{g} }\epsilon^{ab} {\bf X}_a \times {\bf X}_b,
 \ee
 where  $\epsilon^{ab} = - \epsilon^{ba}$ is the 2-dimensional Levi-Civita density ($\epsilon^{12} = +1$), and $g$ is the determinant of the metric $g_{ab}$.

The space vectors $\{ {\bf X}_a , {\bf n} \}$ form a space basis adapted to the surface $\mathbb{S}$. 
In particular,
they satisfy a completeness relationship: given any two space  vectors ${\bf U}$ and ${\bf V}$, one
has that
\begin{equation}
{\bf U} \cdot {\bf V} =
g^{ab} ({\bf U} \cdot {\bf X}_a) ( {\bf V} \cdot {\bf X}_b ) + ({\bf U} \cdot {\bf n}) ( {\bf V} \cdot {\bf n}).
\label{eq:comple}
\end{equation}

The second surface derivatives of the shape functions $\partial_a \partial_b {\bf X}$ can be organized in a tangential and a normal part,
by considering their projections along the space basis using the completeness relationship, as  
\be
\partial_a \partial_b {\bf X} = \left( g^{cd} {\bf X}_d \cdot \partial_a \partial_b {\bf X} \right) {\bf X}_c
+ \left( {\bf n} \cdot \partial_a \partial_b {\bf X} \right) {\bf n}.
\label{eq:proj}
\ee
This simple step leads immediately to the geometrical meaning of the second derivatives.
The tangential projection is the surface Christoffel symbol
\be
\left( g^{cd} {\bf X}_d \cdot \partial_a \partial_b {\bf X}  \right) = \Gamma^c{}_{ab}.
\label{eq:chris}
\ee
The Christoffel symbol can also be given
in terms of the induced metric as
\begin{equation}
\Gamma^c{}_{ab} 
= {1 \over 2} g^{cd } (\partial_a g_{bd} +
\partial_b g_{ad} - \partial_d g_{ab} ).
\end{equation}
In this, perhaps more familiar, form the Christoffel symbol acquires the geometrical meaning of the affine connection
implicit in the surface covariant derivative $\nabla_a = X^\mu_a \partial_\mu$,
compatible with the induced metric $g_{ab}$, such that  for an arbitrary surface vector $v^b$ and a co-vector $u_b$, we have 
\be
\nabla_a v^b =
\partial_a v^b + \Gamma^b{}_{ac} v^c\,,
\quad \quad
\nabla_a u_b =
\partial_a u_b - \Gamma^c{}_{ab} u_c.
\ee
 By compatible it is meant that the induced metric is covariantly constant,
$\nabla_a g_{bc} = 0$.  
Geometrically, the Christoffel symbol  is purely intrinsic, in that
it depends only on the induced metric $g_{ab}$.
In a covariant description of the
surface, the Christoffel symbol  appears only implicitly through the covariant
derivative.  The intrinsic Riemann curvature tensor of the surface $\mathbb{S}$
quantifies the degree of failure of
the surface covariant derivative $\nabla_a$ to commute,
$
(\nabla_a \nabla_b - \nabla_b \nabla_a ) v^c =
{\mathcal R}^c{}_{dab} v^d$, for a surface vector $v^a$.
Contraction of the Riemann tensor gives the
Ricci tensor ${\mathcal R}_{ab} = {\cal R}^c{}_{acb}$,
and the scalar intrinsic curvature  is given by contraction with the
contravariant metric
${\cal R} = g^{ab} {\cal R}_{ab}$.
For a two-dimensional surface, the Riemann tensor is
completely determined by the scalar curvature
\be
{\cal R}_{abcd} = {{\cal R} \over 2} \, ( g_{ac} g_{bd} - g_{ad} g_{bc}),
\label{eq:r2d}
\ee
that implies, in particular,
${\cal R}_{ab} = (1 / 2 ) {\cal R} g_{ab}$. For a two-dimensional surface,
the scalar curvature is twice the Gaussian curvature
${\rm G}$, {\it i.e.}
${\cal R} = 2 {\rm G} $.

The tangential projection of the second derivatives of the space therefore
enters in the covariant derivatives of the tangent vectors as
\begin{equation}
{\bf X}_{ab} = \nabla_a {\bf X}_b = \partial_a {\bf X}_b - \Gamma^c_{ab}\, {\bf X}_c\,, 
\label{eq:xab}
\end{equation}
where we define the shorthand $\xab$ once for all.

Returning to the projections (\ref{eq:proj}), 
the normal projection of the second derivatives of the shape functions identifies the
extrinsic curvature tensor
\begin{equation}
\left( {\bf n} \cdot \partial_a \partial_b {\bf X} \right) = - K_{ab}\,.
\end{equation}
Note that the minus sign is a convention not agreed on by everybody, as not everybody agrees on the letter kappa. The extrinsic curvature
tensor measures the bending of the surface. 
In particular, the eigenvalues $c_1, c_2$  of the
matrix $K_a{}^b =  K_{ac} g^{cb}$ are the principal curvatures of
the surface. The trace  with the inverse metric $g^{ab}$ is the mean extrinsic curvature of the surface
\be
K = g^{ab} K_{ab} = -  g^{ab} {\bf n} \cdot \partial_a \partial_b {\bf X} = -  g^{ab} {\bf n} \cdot {\bf X}_{ab} 
= - {\bf n} \cdot \nabla^2 {\bf X}\,,
\ee
where we have used (\ref{eq:xab}), and $\nabla^2 = g^{ab} \nabla_a \nabla_b$ is the Laplacian. 
With respect
to the principal curvatures, $K = c_1 + c_2 $.

The second surface derivatives of the fields, the shape functions, contain geometric information, whether they like it or not. Their  projections, as given in (\ref{eq:proj}), are then 
expressed geometrically as the sum of an intrinsic and 
an extrinsic part 
\be
\partial_a \partial_b {\bf X} = \Gamma^c{}_{ab}{\bf X}_c
- K_{ab} \, {\bf n}\,.
\ee
This is one of the classical Gauss-Weingarten equations for the surface $\mathbb{S}$ that describe 
the expansion of the surface gradients of the basis $\{ {\bf X}_a , {\bf n} \}$, and that when written in covariant language  are
\begin{eqnarray}
{\bf X}_{ab} &=&  - K_{ab} \; {\bf n}\,, 
\label{eq:gw1}
\\
\nabla_a {\bf n} &=& K_{ab} \; g^{bc} \; {\bf X}_c\,.
\label{eq:gw2}
\end{eqnarray}

In order to have a surface, the intrinsic and extrinsic geometries of $\mathbb{S}$, as described
by the induced metric and the extrinsic curvature tensor, cannot be chosen arbitrarily, but they 
must satisfy the Gauss-Codazzi-Mainardi equations,
that arise as integrability conditions for the Gauss-Weingarten
equations.
 The Gauss-Codazzi-Mainardi  are given by 
\begin{eqnarray}
{\cal R}_{abcd} - K_{ac} K_{bd} + K_{ad} K_{bc} = 0\,,
\label{eq:g1}\\
\nabla_a K_{bc} - \nabla_b K_{ac} = 0\,.
\label{eq:cm1}
\end{eqnarray}
Their contractions with  the contravariant metric $g^{ab}$, that will be used in the text,  are
\begin{eqnarray}
{\cal R}_{ab} -  K K_{ab} +  K_{ac} K_{b}{}^c = 0\,,
\label{eq:g2}
\\
{\cal R} - K^2 + K_{ab} K^{ab} = 0\,,
\label{eq:g3}
\\
\nabla_b K_{a}{}^b - \nabla_a K = 0\,.
\label{eq:cm2}
\end{eqnarray}
For a two-dimensional surface the contracted
Gauss-Codazzi equation (\ref{eq:g3}) contains the same
information as (\ref{eq:g1}). The Gauss theorem (\ref{eq:g3})  says that
Gaussian curvature is given in terms of the principal curvatures
by their product, ${\rm G} = c_1 \; c_2 $. It should be noted that, with respect to the shape functions themselves,
the Gauss-Codazzi-Mainardi equations are only identities, but endowed  with  important geometrical
meaning.

Two equations that are basic for what follows are the expressions for the curvatures
in terms of the derivatives of the shape functions
\ba
K &=& - g^{ab}  ({\bf n} \cdot {\bf X}_{ab} )\,,
\label{eq:kd}
\\
{\cal R} &=& ( g^{ab} g^{cd} - g^{ac} g^{bd} ) ({\bf n} \cdot {\bf X}_{ab} )({\bf n} \cdot {\bf X}_{cd} )\,.
\label{eq:gd}
\ea

\section{Elastic bending energy}

At mesoscopic scales, an homogenous  lipid membrane can be considered as an infinitely thin surface, 
due to the large scale separation between its thickness and its size. The microscopic degrees of freedom
enter in the physical parameters that characterize the membrane, but in a coarse grained description
the conformational degrees of freedom of the membrane are geometrical. 
 Moreover, the membrane 
can be considered as a two-dimensional fluid, as there is no energy cost in  in-plane displacements of the
constituent lipid molecules. The infinitesimal tangential displacements are represented by
reparametrizations  of the surface that represents the membrane.  A suitable energy density for a homogenous
 membrane therefore must be invariant under reparametrizations, and constructed from the surface
geometric scalars, in short a geometric model \cite{Kamien-rmp,Deserno15}. 

As opposed to tension dominated systems, like soap bubbles, in lipid membranes the important mode of deformation
is bending, and the relevant geometric scalars are constructed from the curvatures of the surface. 
The elastic bending energy that underpins   
the current theoretical understanding of the morphology and of the shape transformations of lipid membranes is given by the classic Canham-Helfrich bending elasticity model \cite{Can,Hel,Evans} 
\be
F  [ X ] =  \int_{\mathbb{S}}  dA \left[ \kappa  \, (K - K_0)^2 + \overline{\kappa} \, {\cal R}  + \sigma \right]\,.
\label{eq:elas1}
\ee
where the integral is over the surface $\mathbb{S}$. 
The free energy depends on the shape functions through the 
mean curvature $K$, the intrinsic curvature ${\cal R}$, and the area element $dA = \sqrt{g} d^2 \xi$, and it
contains four phenomenological parameters. The parameter $\kappa$ is the bending rigidity, and $\overline\kappa$ is the Gaussian bending rigidity, with dimensions of energy. 
The parameter $\sigma$ can be interpreted as a surface tension, or as a chemical potential that takes into account
the insolubility of the membrane, {\it i.e.} the fact that the number of lipid molecules remains  constant on the time scale of interest. 
The parameter $K_0$ is the spontaneous curvature, with dimensions of inverse length, and it
describes  a possible asymmetry of the membrane. Other more realistic curvature models have been proposed to take into account the bilayer architecture, notably the bilayer couple model
\cite{Svetina89}
that implements as a constraint  the area difference functional 
\be
F_{(M)} [X] = \beta \int_{\mathbb{S}}  dA \, K\,.
\label{eq:fmean}
\ee
where the parameter  $\beta$ is proportional to the spontaneous curvature. A different refinement is  the area-difference-elasticity model,
that involves a non-local constraint, see \cite{Boz92,Area}, and in particular the recent review \cite{SZ2014}.

The Canham-Helfrich model  contains two geometric invariants well known in differential geometry.
The first is the integral over the two-dimensional surface of the intrinsic scalar curvature. If the surface has no boundary,
it is a topological invariant, because of the Gauss-Bonnet theorem \cite{doCarmo}
\begin{equation}
\int_{\mathbb{S}} dA \, {\cal R} = 8 \pi  (1 - {\rm g})\,,
\label{eq:gb}
\end{equation}
where ${\rm g}$ is the genus of the surface. For a closed membrane with fixed genus, the Gaussian bending
rigidity has no effect on the equilibrium condition.
The second geometric invariant is the integral over the surface of the squared mean curvature, or bending energy, that is known as the Willmore functional \cite{Willmore}.
The Willmore functional is not only scale invariant, but also invariant under conformal transformations in space, and in differential geometry it
plays a fundamental  role in the combination of minimal surface theory and conformal invariance \cite{BGS,RiviereC}.

The Canham-Helfrich effective energy  and its theoretical refinements  
have been the subject of extensive  reviews \cite{SL1,Seifert,Turecent,Deserno15}.
For a recent review of the remarkable effectiveness of the guidance offered by   this
phenomenological theoretical model in the laboratory,  see \cite{Bass}.

As a geometric model, the Canham-Helfrich  functional can be considered as an  especially  interesting case of a larger
class of reparametrization invariant geometric models that describe physical degrees of freedom localized on a two-dimensional
sub-manifold The geometric models are obtained 
by a  natural expansion in terms of derivatives of the shape
 functions, or equivalently in a smallness parameter given by the ratio of the thickness 
with respect to the inverse curvature,  with a general free energy of  the form 
\be
F [ X ] = \int_{\mathbb{S}} dA
\, f  ( g_{ab}, K_{ab}, \dots)\,,
\ee
where the energy density $f  ( g_{ab}, K_{ab}, \dots) $ is given by geometric scalars of the intrinsic
and the extrinsic geometries of the surface. 
Rather than using composite field variables, like  the surface metric and the extrinsic curvature 
as a means to construct geometric scalars, as it is common usage,
however, one can use
directly the shape functions and their derivatives as field variables, and consider an equivalent  generic free energy
 \begin{equation} 
F [ X ] = \int_{\mathbb{S}} dA  f ( \partial_a {\bf X}, \partial_a \partial_b {\bf X}, \dots  )\,.
\end{equation}
This is the approach taken in this paper, where the emphasis is on the shape functions,  and the energy density is allowed to be even more general,
including a dependance on the shape functions themselves, and on the surface coordinates, breaking
the reparametrization invariance symmetry, to include in a systematic way external forces, and
inhomogeneities of the constituent lipids.

The use of the shape functions as field variables is  motivated also by the fact that, in field theory, 
membranes and interfaces are described using  the height representation, where
the surface is represented by a height function $h(x,y)$ over a reference planar configuration, and one considers
a general energy  
\begin{equation} 
F [ h ] = \int_{\mathbb{S}} dA \,  f ( \partial_a h , \partial_a \partial_b h,\dots )\,.
\end{equation}
When the surface is almost planar, $|\partial h | \ll 1$,  the energy density simplifies to a tractable level. For example,
 the Canham-Helfrich model (\ref{eq:elas1}) in this linearized 
approximation becomes quadratic in the fields
\be
F  [ X ] =  \int_{\mathbb{S}}  dx dy  \left[ \kappa  \, (\nabla^2 h )^2 + \sigma (\nabla h)^2
\right]\,,
\label{eq:elash}
\ee
where we have set $K_0 = 0$ and $\overline\kappa = 0$, for the sake of simplicity.
This approximation is appropriate when one is studying short wavelength thermal fluctuations of the
surface \cite{Jerusalem,Safram}. 

 \section{First variation: shape equation, linear and angular stress tensors}
 
 In this section, we consider the minimization and the symmetries of the  elastic bending free energy, using a
 manifestly covariant variational approach.   
 This variational approach  treats the geometric model as a problem of classical field theory, in the framework of the
 familiar continuum Lagrangian formulation \cite{Gold,FW}, albeit extended to higher derivatives \cite{Rund,deleon}.  The higher-derivative field theoretical formalism can be compared to an acceleration dependent Lagrangian
formulation for a classical point particle. 
 Of course, this classical mechanics analogy  is only formal, yet it provides useful sign posts, and foremost a connection to a familiar language. 

 The classical fields are
 the shape functions that describe the membrane. In the Lagrangian formulation, the shape functions ${\bf X} (\xi^a) $ are  the generalized coordinates. The first derivatives
 of the shape functions, the two tangent vectors $\xa$, are the rate of change of the generalized coordinates, or `generalized velocities'. Since the model depends
 on the curvature, it involves the second derivatives of the fields. We choose  to use the covariant second 
 derivatives of the shape functions ${\bf X}_{ab} = \nabla_a \nabla_b {\bf X}$ as rate of change of the generalized velocities, or  `generalized accelerations', rather than  the non-covariant option
 $\partial_a \partial_b {\bf X}$. This choice can be made precise resorting to the jet bundle formulation of 
 higher order variational calculus and its symplectic geometry \cite{jet}, to show that there is no loss of generality, but for the sake 
 of simplicity we will just assume it. Note that, by the definition of the covariant derivative, the generalized velocities and the generalized accelerations
 are perpendicular, $ \xa \cdot {\bf X}_{bc} = 0$, but in the Lagrangian formulation they are treated as independent. 
From a strict field theoretical point  of view, this orthogonality condition should be added as a non-holonomic constraint,
but doing so unnecessarily complicates the formulation, and ultimately has no effect.
One crucial feature of the Lagrangian formulation, often described as a mere convenience but in fact
essential,  is the freedom of choosing suitable  local coordinates, where by suitable it is meant that they
must span appropriately the configurational phase space.
 The extended higher-derivative Lagrangian phase space we adopt is given by the shape functions and its covariant derivatives,  $\{ {\bf X}, \xa, \xab \}$.
  The surface covariant derivative $\nabla_a$ measures the rate of change, just like  the time derivative in
 the classical mechanics of a point-like particle, in the spirit if not the letter of a two-dimensional  relativistic field theory,
 after a Wick rotation, of course.

 The role of the Lagrangian is played by the 
  energy density, a scalar density of weight one. 
 The most general energy density with a dependance  at most on two derivatives of the shape functions 
 is
 \be
 \mf = \mf ({\bf X}, \xa, \xab, \xi^a)\,.
 \label{eq:general}
 \ee
  The word density is used both in its physical, dimensional, meaning, and in its 
 mathematical meaning as a densitized function. Concretely, the energy density can be written as 
  $\mf  = \sqrt{g} \, f $,  where $f$ is a scalar function. In general,  it is preferred to use the scalar function $f$ to represent the energy density, as in the previous section when introducing the elastic bending energy.  In any case the square root of the metric is dimensionless, and it is only a matter
 of convenience,
but the choice of a densitized function is natural from a Lagrangian viewpoint, and it will show to be not only advantageous,
but also almost mandatory.  
 To establish the Lagrangian formulation, in this section  we specialize our considerations to a
reparametrization invariant  energy density of the form
 \be
 \mf = \mf (\xa, \xab )\,.
 \ee
with a dependance  at most on two derivatives of the shape functions, and for the moment without
a dependance on the shape functions themselves, or on the surface coordinates. 
In the classical mechanics analogy, we are focusing for the moment  our attention on the `kinetic' part
of the Lagrangian. The general case (\ref{eq:general})  is considered in the following section.

The role of the action is taken by the free energy functional, the integral of the energy density over the surface,
\be
F [ X ] = \int_{\mathbb{S}}  {\mathcal F} ( {\bf X}_a , {\bf X}_{ab})\,.
\label{eq:fx1}
\ee 
Here and in the following we absorb the differential $d^2\xi$ in the integral sign.
 The symmetries of this free energy are reparametrization invariance and invariance
 under rigid motions in space. 

To illustrate the formalism, we will use as representative examples the energy density for a soap film, proportional 
to the area, the bending energy density, the mean curvature density, and the Gaussian bending energy density, given respectively by  
\be
\mf_{(S)} = \sigma \;  \sqrt{g}\,,
\label{eq:soap}
 \ee
\be
\mf_{(B)}  = \kappa   \sqrt{g} \; K^2\,,
\label{eq:bending}
\ee
\be
\mf_{(M)} = \beta \, \sqrt{g} K\,,
\label{eq:mean}
 \ee
  \be
\mf_{(G)} = \overline{\kappa} \, \sqrt{g} \; {\cal R} = \overline{\kappa} \, \sqrt{g} \,\left( K^2 - K^{ab} K_{ab} \right)\,,
\label{eq:gaussian}
\ee
where $\beta$ is a material parameter proportional to the spontaneous curvature.
The energy density for the  bilayer-couple  Canham-Helfrich model can be written as their sum
\be
\mf_{(CH)} = \mf_{(B)} +  \mf_{(G)} + \mf_{(M)} + \mf_{(S)}\,.
\label{eq:CH}
\ee

 \subsection{First variation}
 
Consider an infinitesimal variation of the shape functions
\[
{\bf X} (\xi^a) \to {\bf X} (\xi^a) + \delta {\bf X} (\xi^a) = {\bf X} (\xi^a) +  {\bf W} (\xi^a)\,, 
\] 
where the symbol ${\bf W} = \delta {\bf X}$ is introduced for the first variation of the shape functions, 
to simplify the notation in what follows.
In order to maintain manifest covariance thoughout,  
its surface covariant derivatives are defined as
\[
{\bf W}_a = \nabla_a {\bf W} = \partial_a {\bf W} \,, \quad \quad {\bf W}_{ab} = \nabla_a \nabla_b {\bf W}\,.
\]

The first variation of the energy density   with respect to variations of the shape functions is
\be
\delta  \mf ({\bf X}_a , {\bf X}_{ab})  = {\partial \mf \over \partial {\bf X}_a } \cdot  \delta {\bf X}_a + {\partial \mf \over \partial {\bf X}_{ab } } \cdot \delta {\bf X}_{ab}\,.
\label{eq:ff1}
\ee

Manifest covariance  suggests  to write  the variation  in covariant form as a gradient along the variation vector ${\bf W}$ as $ \delta 
=  W^\mu \partial_\mu $. 
The mathematical variation coincides with an infinitesimal
deformation of the membrane. The variation is conserved along the surface, and, geometrically, 
this translates to assuming that the Lie derivative of the variation along the tangent vectors
vanishes, or $[\delta , X_a ] = [\delta , \nabla_a ] = 0$.  From this simple assumption, it follows that 
\ba
\delta {\bf X}_a &=& [ \delta , \nabla_a ] \, {\bf X} +  \nabla_a \, \delta {\bf X} = {\bf W}_a\,,
\label{eq:w1}
\\
\delta {\bf X}_{ab}  &=& [ \delta , \nabla_a ] \partial_b {\bf X} + \nabla_a \delta \, \partial_b {\bf X} 
 = \nabla_a \nabla_b \delta {\bf X} = {\bf W}_{ab}\,,
\label{eq:w2}
 \ea
using the vanishing of the commutator. The use of a covariant variation permits to avoid having to deal with the
variation of the affine connection implicit in the covariant derivative, at least at first order.   In a curved 
ambient space, however, the commutator  would not vanish, and it would contribute a term involving the ambient curvature,
as is the case for a relativistic extended object, or brane, in a curved background spacetime, see {\it e.g.} \cite{CarterBM,defo}.  
 The covariant variational derivative is of common usage in relativistic field theory, for example 
 in the derivation of the geodesic deviation equation in General Relativity \cite{Wald,bz77b}.

The first order variation of the energy density  (\ref{eq:ff1}) takes then the form
\be
\delta  \mf ({\bf X}_a , {\bf X}_{ab})  = {\partial \mf \over \partial {\bf X}_a } \cdot {\bf W}_a + {\partial \mf \over \partial {\bf X}_{ab } } \cdot {\bf W}_{ab}\,.
\label{eq:fo1}
\ee
This variational differential plays a fundamental role in the Lagrangian formulation.
At first order, its covariance allows never to contend with  the affine
connection, except implicitly in the covariant derivative. The same variation of the energy density can also be obtained using the language of  differential
forms and exterior calculus \cite{Grif}, but this requires  the necessary mathematics investment. 
 This covariant variation of the energy  density  has the same structure 
as the one that appears in the variational approach in the height representation,
but of course without the need for a planar reference configuration.

With the help of the variational differential, 
the first variation of the free energy is therefore simply
\begin{equation}
\delta F [ X ] = \int_{\mathbb{S}}  \delta \mf =\int_{\mathbb{S}} \left[ \left({ \partial \mf \over \partial {\bf X}_a } \right) \cdot  {\bf W}_a
+  \left( {\partial \mf \over \partial {\bf X}_{ab} } \right) \cdot  {\bf W}_{ab}  \right]\,.
\end{equation}
The higher-derivative  Lagrangian formulation follows naturally. 
Integrating by parts twice gives immediately
\be
 \delta F [ X ] = \int_{\mathbb{S}}\,  \left[  \boldsymbol{\cal E} (\mf)  \cdot {\bf W}  
+
\nabla_a {\cal Q}^a (W) \right]\,.
\label{eq:varf1}
\ee 
The Euler-Lagrange derivative of the energy density is
\be
 \boldsymbol{\cal E} (\mf) =  - \nabla_a \left({ \partial \mf \over \partial {\bf X}_a } \right) 
+  \nabla_a \nabla_b \left( {\partial \mf \over \partial {\bf X}_{ab} } \right)\,.
\label{eq:el}
\ee
The divergence form of the Euler-Lagrange derivative is only a direct consequence of the
restriction to a derivative dependent energy density, and the usefulness of the use of a densitized energy
density should be apparent.

The Noether current is given by the surface vector density 
\be
{\cal Q}^a (W)  = \left[  {\partial \mf \over \partial {\bf X}_a }  - \nabla_b \left( {\partial \mf \over \partial {\bf X}_{ab} } \right) \right]  
\cdot {\bf W}
+ {\partial \mf \over \partial {\bf X}_{ab} } \cdot {\bf W}_b\,.
\label{eq:emmy}
\ee
When integrated over the surface, the divergence  of the Noether current gives a boundary term, that vanishes for a closed surface.  If the membrane is open, natural boundary conditions are dictated by this boundary term, see {\it e.g.} \cite{Nitsche}. The covariance of the variational derivative  is what produces
the simple structure of this boundary term. 
 The Noether current is not uniquely determined. There is a Lagrangian `gauge freedom' of adding a gradient
 that leaves the total divergence unchanged 
  \be
{\cal Q}^a  \to {\cal Q}'^a  = {\cal Q}^a + 
\epsilon^{ab} \partial_b \tilde{g}\,,
\label{eq:ambi}
\ee
where $\tilde{g}$ is an arbitrary scalar density, and $\epsilon^{ab} = - \epsilon^{ba}$ is the surface Levi-Civita symbol. This ambiguity is
the same as the familiar freedom of adding a total derivative to the Lagrangian in classical mechanics. 
In the context of lipid membranes, it has been studied in detail  in \cite{LM}.

In  order to compare with  other variational approaches to the minimization of the free energy, the variation of the
shape functions can be 
decomposed in a standard way in its normal and tangential parts as
\be
{\bf W} = W_\perp {\bf n} + W^a {\bf X}_a\,.
\ee
Then, since the tangential variation can be identified with  a reparametrization of the surface, the tangential variation of the energy 
density contributes only a total derivative, $\delta_\parallel
\mf = \nabla_a (W^a \mf)$.  Therefore, the first variation (\ref{eq:varf1}) takes the form
\be
\delta F [X] = \int_{\mathbb{S}} \left\{ \boldsymbol{\cal E} (\mf)  \cdot {\bf n}  \,W_\perp + 
\nabla_a \left[  {\cal Q}^a (W_\perp)  + W^a {\cal F} \right] \right\}\,.
\ee
 If one is  interested mainly in the determination of  the equilibrium condition
 it is natural to focus on normal variations from the outset, and the tangential variations can be safely neglected as unphysical.  Only the normal part of the Euler-Lagrange derivative  enters in the equilibrium condition.  As a consequence of reparametrization invariance,
the tangential part of the Euler-Lagrange derivative vanishes identically
\be
\boldsymbol{\cal E} (\mf)  \cdot {\bf X}_a = 0\,.
\label{eq:elt}
\ee
For our purposes, however, a shortcoming of the normal/tangential decomposition is that it obscures the structure of the Lagrangian formulation. 
Moreover, the decomposition becomes less useful whenever the symmetry of  reparametrization invariance is broken,
as in the case of heterogenous membranes, or in the presence of boundaries, or domains. 

With respect to approaches that model the membrane using liquid crystal elasticity theory \cite{Napolieq}, it should be 
noted that the second covariant derivative $\xab$ plays the role of a director field, as in the Franck elasticity energy, for the constituent lipid molecules, aligned along the normal to the surface. A possible tilt degree of freedom of the molecules can always be included as a second step as a surface vector field \cite{PN}.
In this regard, it is interesting to note the indirect appearance  of a discrete sub-structure in a continuum model.

\subsection{Shape equation}

The vanishing of the first variation of the free energy defines equilibrium,
\be
\delta F [ X ] = 0\,.
\ee
For a closed membrane, the boundary term vanishes. For a membrane with a non-empty boundary
natural boundary conditions need to be imposed to make it vanish \cite{Nitsche}.
Therefore, the vanishing of the first variation
 (\ref{eq:varf1})  reduces to
 \be
 \delta F [ X ] = \int_{\mathbb{S}} \, \boldsymbol{\cal E} (\mf)  \cdot {\bf W}\,.
 \label{eq:var0}
\ee
Since the variation ${\bf W}$ is arbitrary, by the fundamental lemma of the calculus of variations,  the vanishing of the first variation implies  the Euler-Lagrange 
equations that determine the equilibrium configurations of the membrane
\be
\boldsymbol{\cal E} (\mf) =  - \nabla_a \left({ \partial \mf \over \partial {\bf X}_a } \right) 
+  \nabla_a \nabla_b \left( {\partial \mf \over \partial {\bf X}_{ab} } \right) = 0\,.
 \label{eq:ello}
 \ee
 The equilibrium condition equation takes automatically a divergence-free form. This
 fact will be discussed in the next sub-section, when we introduce the linear stress tensor. 
 The Euler-Lagrange derivative of the energy density  represents the force ${\bf F}$ per unit area acting on the membrane 
 due to bending elasticity. It can be understood as the elastic force exerted on a patch of the membrane by the rest 
 of the membrane. Its direction is chosen to be from the boundary of the patch into the rest of the membrane, according to 
 a standard, but not universally accepted, convention in elasticity theory  \cite{LL,Deserno15}. 
 As a consequence of reparametrization invariance, the tangential projections are satisfied identically, see  the identity (\ref{eq:elt}),
and the force density is normal to the surface. For an acceleration dependent classical particle, the Euler-Lagrange 
derivative has the same structure.

 The equilibrium condition (\ref{eq:ello}) for the Canham-Helfrich model has come to be known as
 `shape equation'. We  use this expression for the equilibrium condition of any
 geometric model. The simplicity of the mathematical structure of the Euler-Lagrange equations (\ref{eq:ello}) makes
 the derivation of the shape equation for any geometric model a straightforward exercise.
 Let us consider some examples of shape equations, to illustrate the formalism.
  For a soap film, the energy density (\ref{eq:soap})  is proportional to the area, and as it depends only on the first derivatives
of the shape functions, all we need is one partial derivative
\be
{ \partial \mf_{(S)} \over \partial {\bf X}_a } = \sigma \sqrt{g} g^{ab} {\bf X}_b\,.
\label{eq:pds}
\ee
It follows  that the Euler-Lagrange derivative, or the force density, is
\be
\boldsymbol{\cal E} (\mf_{(S)}) =   - \nabla_a \left( { \partial \mf_{(S)} \over \partial {\bf X}_a } \right) = 
 - \sigma \nabla_a \left( \sqrt{g} g^{ab} {\bf X}_b \right) = - \sigma \sqrt{g} \, \nabla^2 {\bf X}\,,
 \label{eq:elsp}
 \ee 
 where $\nabla^2 = g^{ab} \nabla_a \nabla_b = (1 / \sqrt{g} )\partial_a (\sqrt{g} g^{ab} \partial_b )$ is the Laplacian. 
 Geometrically, using (\ref{eq:gw1}), this gives that the shape equation, written in covariant form, for a soap film is the vanishing of the mean curvature,
 \be
 \boldsymbol{\cal E} (\mf_{(S)}) = \sigma \sqrt{g} \, K \, {\bf n} = 0\,,
 \ee
and one recognizes  the equation for a minimal surface.
Note that the force density in this case  is automatically normal to the surface.
Compared to lipid membranes, soap films are simpler only in the sense that their energy density does not depend on higher derivatives through
the curvature, but it should be appreciated that,
in mathematics, minimal surfaces constitute 
a subject area of its own \cite{FT,minimal}. 

For the derivation of the shape equation for the bending energy density (\ref{eq:bending}), we need two partial derivatives 
\ba
  {\partial \mf_{(B)} \over \partial {\bf X}_a } &=& \kappa \sqrt{g} \, K \, \left( K g^{ab} - 4   K^{ab} \right) {\bf X}_b\,,
\label{eq:pdb1}
\\
{\partial \mf_{(B)} \over \partial {\bf X}_{ab} } &=& - 2\kappa \sqrt{g} K g^{ab} {\bf n}\,.  
\label{eq:pdb2}
 \ea 
Simply plugging in the general expression (\ref{eq:el}), one finds that the Euler-Lagrange derivative is   given by
  \ba
 \boldsymbol{\cal E} (\mf_{(B)}) &=& \kappa \nabla_a \left[  \sqrt{g}  K \left( 4 K^{ab} -  K  g^{ab} \right){\bf X}_b \right]
 - 2  \nabla_a \nabla_b \left( \sqrt{g} K g^{ab}  {\bf n} \right)  \nn \\
&=& \kappa \sqrt{g}\; \nabla_a   \left[ K ( 2 K^{ab}  -  K g ^{ab} ) {\bf X}_b - 2 g^{ab} (\nabla_b  K )\, {\bf n}  \right]   \nn 
 \\
  &=& \kappa \sqrt{g} \left[ 2 K \left(\nabla_a K^{ab} - \nabla^b K \right)  {\bf X}_b 
 +\left( - 2\nabla^2 K + K^3 - 2K K_{ab} K^{ab} \right)  {\bf n} \right]\,, 
 \label{eq:shape1}
 \ea 
 where the Gauss-Weingarten equations (\ref{eq:gw1}), (\ref{eq:gw2}) have been used.
  In the  third line,  one sees that the tangential components vanish, as expected, because of the contracted Codazzi-Mainardi equations (\ref{eq:cm2}), and the identity (\ref{eq:elt}) is satisfied. The relation of the vanishing of the tangential components to the Codazzi-Mainardi
 equations is examined in \cite{LM}.  With the help of the Gauss theorem
 (\ref{eq:g3}), the normal component gives the well known shape equation for the bending energy \cite{Jenkins,Hel87,Hel89} 
\be
\boldsymbol{\cal E} (\mf_{(B)}) = \kappa \sqrt{g} \left[ - 2\nabla^2 K - K^3 +  2 K {\cal R} \right] \,{\bf n} = 0\,.
\label{eq:shape2}
 \ee
 This shape   equation  is a  fourth order non-linear partial differential equation, as expected from classical elasticity theory, and it is  considerably more difficult to solve than the shape equation for a soap film (\ref{eq:elsp}). 
 In mathematics, the study of the shape equation for Willmore surfaces is the subject of  active current 
investigation, and it has reached the status of a research area, see {\it e.g.} the recent 
accomplishments  of \cite{Marques,BR}.

 Besides the bending energy, the mean curvature energy density (\ref{eq:mean})  is also needed  in the Canham-Helfrich model for the inclusion of the spontaneous curvature, but it is important to note  the fact that the  linearity in the second derivatives  means that the mean curvature energy density
is not a higher-derivative field theory. For a classical particle, see ex. 12 of Ch. 2 in \cite{Gold}.   We have
the partial derivatives
\ba
 {\partial \mf_{(M)} \over \partial {\bf X}_a } &=& \beta \sqrt{g} \, \left( K g^{ab} - 2  K^{ab} \right) {\bf X}_b\,,
\\
{\partial \mf_{(M)} \over \partial {\bf X}_{ab} } &=& - \beta \sqrt{g}  g^{ab} {\bf n}\,,  
 \ea 
and from the general expression (\ref{eq:ello}), the Euler-Lagrange derivative or the force density is 
\ba
  \boldsymbol{\cal E} (\mf_{(M)})  &=& \beta \, \nabla_a \left[  \sqrt{g} \, \left( K g^{ab} - 2  K^{ab} \right) {\bf X}_b - \nabla_b ( \sqrt{g} g^{ab} {\bf n} )\right] \nn \\ 
&=&  \beta \sqrt{g} \left[ (\nabla_a K^{ab} - \nabla^b K ) {\bf X}_b + (K^2 - K_{ab} K^{ab} ) {\bf n}  \right] \nn \\ 
 &=& \beta \sqrt{g} \, {\cal R} \, {\bf n}\,, 
\ea
where, again, the contracted Codazzi-Mainardi equations (\ref{eq:cm2}) make the tangential part vanish,  and the Gauss equation (\ref{eq:g3}) has been used to obtain the last equality. The force density is normal, and its component  depends
only on the intrinsic geometry of the surface, through the intrinsic curvature.

For the Gaussian bending energy density (\ref{eq:gaussian}), we expect a vacuous shape equation,  
since its integral over a two-dimensional closed surface  is a topological invariant, by the Gauss-Bonnet theorem (\ref{eq:gb}). 
To confirm it within this approach, consider that, using the definition (\ref{eq:gd}), the
partial derivatives are 
\ba
{\partial \mf_{(G)} \over \partial {\bf X}_a } &=& - \overline\kappa  \, \sqrt{g}  \,  {\cal R}  \,g^{ab} {\bf X}_b\,, \\
{\partial \mf_{(G)} \over \partial {\bf X}_{ab} } &=&  2 \overline\kappa \sqrt{g} \left( K^{ab}  - K g^{ab} \right) {\bf n} \,. 
 \ea 
 It follows that the Euler-Lagrange derivative is
\ba
\boldsymbol{\cal E} (\mf_{(G)}) &=& \overline\kappa \sqrt{g} \nabla_a \left\{ {\cal R}  g^{ab} {\bf X}_b  + 
2 \nabla_b  \left[ \left( K^{ab}  - K g^{ab} \right) {\bf n}  \right] \right\} \nn \\
&=& \overline\kappa \sqrt{g} \nabla_a  \left\{{\cal R}  g^{ab} {\bf X}_b  + 
 2 (\nabla_b  K^{ab}  - \nabla^a  K )  {\bf n}  - 2 (K^{ac} K_c{}^b - K K^{bc} ) {\bf X}_b \right\} 
\nn \\
 &=& \overline\kappa \sqrt{g} \nabla_a \left[ ({\cal R} g^{ab} - 2 {\cal R}^{ab} ) {\bf X}_b \right] = 0\,,
\label{eq:elg}
\ea
where we have used, yet again, the contracted Codazzi-Mainardi equation (\ref{eq:cm2}), and the Gauss theorem (\ref{eq:g2}) to obtain the third line.
The last equality follows from the two-dimensional geometrical fact ${\cal R}_{ab} = (1/2) {\cal R} g_{ab}$,
see (\ref{eq:r2d}).  
As expected, the Euler-Lagrange derivative vanishes identically.
Note that, so far, this is the only instance where the two-dimensionality of the surface has been used.
In particular, this observation implies that, mathematically, the Lagrangian formulation applies also
to geometric models for hypersurfaces of arbitrary dimension.

The shape equation for the Canham-Helfrich geometric model (\ref{eq:CH}) can be readily obtained by summing
up all the single contributions as
\be
\boldsymbol{\cal E} (\mf_{(CH)}) = \sqrt{g} \left[ \kappa \left( - 2\nabla^2 K - K^3 +  2 K {\cal R} \right)
+\beta   {\cal R} + \sigma  K \right] \, {\bf n} = 0\,. 
\label{eq:chs}
\ee
 There has been extensive work in the exploration of the
 space of  solution of this shape equation, both analytically and numerically, see the reviews \cite{SL1,Seifert,Turecent}. On the analytic side, a natural approach is the reduction to axisymmetric
 configurations, or surfaces of revolution, motivated both  by the simplification it implies, and  by its relevance
 in biological systems that often prefer symmetric configurations.

 Admittedly, these derivations of the shape equation for a given energy density can be considered  a bit tedious, but no more than any exercise in classical mechanics using the Lagrangian formulation. In comparison to approaches 
 that exploit the normal/tangential decomposition  using differential geometry in the variation of the free energy, lists
 of geometrical variations are traded for partial derivatives.  

 \subsection{Linear stress tensor}

 The free energy is invariant under constant translations in space.
 For the derivation of the conserved linear stress tensor, it is natural to use Noether's theorem
 \cite{stress02,stress04}, and especially so in a Lagrangian formulation.
  Let us consider an
 infinitesimal translation 
 \[
 {\bf W} = \delta_T {\bf X} = {\bf a} = \mbox{const.}
 \]
 The Noether current
${\cal Q}^a (W) $, as defined in
(\ref{eq:emmy}), for a translation specializes to
\be
{\cal Q}^a (a) =  \left[  {\partial \mf \over \partial {\bf X}_a }  - \nabla_b \left( {\partial \mf \over \partial {\bf X}_{ab} } \right) \right]  
\cdot {\bf a} = - \tilde{\bf f}^a \cdot {\bf a}\,,
\ee  
 where, according to Noether's theorem, 
  the conserved  linear stress tensor that follows from translational 
invariance can be simply read off as
\be
\tilde{\bf f}^a =  - {\partial \mf \over \partial {\bf X}_a }  + \nabla_b \left( {\partial \mf \over \partial {\bf X}_{ab} } \right)\,.
\label{eq:stressf}
\ee
A tilde is used to signal that the linear stress tensor is a vector density of weight one, $\tilde{\bf f}^a = \sqrt{g} \, {\bf f}^a$,
where ${\bf f}^a$ is the stress tensor  to be found in the literature \cite{Deserno15,stress02,Fournier}.
The simplicity and the practical usefulness of this general expression for the linear stress tensor needs
to be emphasized.

In the classical mechanics analogy, the linear stress tensor corresponds to minus the canonical  linear momentum
conjugate to the shape functions. The minus sign comes from the convention adopted  about the
direction of the force density as pointing inside the membrane from a boundary. The densitization
of the stress tensor can be seen as appropriate, since the linear momentum is  geometrically a one-form, as is the force,  facts usually
not sufficiently emphasized in standard classical mechanics textbooks. 
Defining the  bending momentum, 
\be
\tilde{\bf f}^{ab} = - {\partial \mf \over \partial {\bf X}_{ab} }\,, 
\label{eq:fab}
\ee
the classical mechanics correspondence extends to the canonical higher linear momentum conjugate to the  tangent vectors, the generalized velocities, again up to a sign. In the classical mechanics analogy, the 
higher-derivative Lagrangian phase space is thus given by the two  conjugate 
pairs $\{ {\bf X} , \tilde{\bf f}^a \}$ and $\{ \xa , \tilde{\bf f}^{ab} \} $, and it can be considered the formal basis of the whole Lagrangian formulation. The identification of the phase space variables can be used in the construction of a covariant
Hamiltonian formulation, to complement the canonical Hamiltonian formulation of \cite{Ham1}, and this will be
considered elsewhere.

Simple comparison of the Euler-Lagrange derivative of the energy density (\ref{eq:el}) with the divergence of the stress tensor (\ref{eq:stressf})  shows that
they are equal
\be
 \boldsymbol{\cal E} (\mf)  =  \nabla_a \tilde{\bf f}^a\,. 
 \label{eq:der1f}
\ee
Therefore at equilibrium, when the shape equation is satisfied,
the linear stress tensor is conserved, or divergence-free,
\be
{\bf F} = \nabla_a \tilde{\bf f}^a = 0\,,
\ee
where ${\bf F}$ denotes the force density acting on the membrane.
That the equilibrium condition can be written as a conservation law should not come as a surprise.
It is just a consequence of the absence of external forces, together with the fact that the
membrane is an extended object. In the classical mechanics analogy, it corresponds 
simply to $\dot{\bf p} = 0$, as emphasized in  \cite{Deserno15}. However, one should also admit 
that from a  cursory look at the shape equation for the Canham-Helfrich model (\ref{eq:chs}), the fact that 
it can be expressed as a conservation law is not quite so obvious; it is only after the role 
of the linear stress tensor has been recognized. It should be also be  added that mathematicians attach 
a very high value to writing  a non-linear partial differential equation in divergence form.

The simplicity of the general expression (\ref{eq:stressf}) makes  the derivation of the linear stress tensor 
for a given energy density straightforward. In fact, in retrospect, one should have considered the linear stress
tensor first,  before the shape equation, but we decided to follow tradition.  
To illustrate the formalism, let us consider some examples.
For a soap film, described by the energy density (\ref{eq:soap}), the linear stress tensor is isotropic and tangential,
\be
\tilde{\bf f}^a_{(S)}  =  -  {\partial \mf_{(S)} \over \partial {\bf X}_a } = -  \sigma \, \sqrt{g} \, g^{ab} {\bf X}_b\,. 
\label{eq:fsp}
\ee
Its divergence gives a force density proportional to the mean curvature
\be
{\bf F}_{(S)}  = \nabla_a \tilde{\bf f}^a_{(S)} = \sigma \sqrt{g} K \, {\bf n}\,,
\ee
that, with our convention for the sign of the mean curvature, makes a soap bubble collapse if pierced.

For the bending elastic energy (\ref{eq:bending}), the linear stress tensor is
\be
\tilde{\bf f}_{(B)}^a = \kappa \left[ \sqrt{g} K ( 4 K^{ab} - K g^{ab} )  {\bf X}_b - 2 \nabla_b ( \sqrt{g} g^{ab} K {\bf n} ) \right]\,,
\label{eq:stressb1}
\ee
as one can calculate directly using the partial derivatives (\ref{eq:pdb1}), ({\ref{eq:pdb2})  in the general expression
(\ref{eq:stressf}), or  alternatively read off from the second line of (\ref{eq:shape1}).
Using the Gauss-Weingarten  equation (\ref{eq:gw2}), the stress tensor can be projected in its tangential 
and normal components as \cite{stress02,stress04,Fournier}
\be
\tilde{\bf f}_{(B)}^a =   \kappa \sqrt{g} \left[ K ( 2 K^{ab}  -  K g ^{ab} )  {\bf X}_b - 2 g^{ab} (\nabla_b  K ) {\bf n}  \right], 
\label{eq:fben}
\ee
Note that the tangential projection is traceless in two dimensions, because of the conformal invariance
of the bending energy.
In this form, the linear stress tensor has been shown to be an extremely useful tool in the mechanics
of lipid membranes. We refer the reader to the review by Deserno \cite{Deserno15} for a 
thorough discussion and a careful analysis of its physical meaning, together with applications
to concrete physical systems.

For the  mean curvature energy density, the linear stress tensor is given by
\be
\tilde{\bf f}^a_{(M)} = \beta   \sqrt{g} \, \left(2 K^{ab} -   K g^{ab} \right) {\bf X}_b  - \nabla_b \left(\beta \sqrt{g}  g^{ab} {\bf n} \right)\,, 
\label{eq:fmean1}
\ee
or, using (\ref{eq:gw2}), it turns out to be purely tangential,
\be
\tilde{\bf f}^a_{(M)} =  \beta  \sqrt{g} \, \left( K^{ab} - K g^{ab} \right) {\bf X}_b\,,
\label{eq:fmean2}
\ee
as expected from the linearity in second derivatives of the energy density. 

For the Gaussian bending energy, one finds that the linear stress tensor vanishes
identically
\be
\tilde{\bf f}^a_{(G)} = \overline\kappa \sqrt{g} ({\cal R} g^{ab} - 2 {\cal R}^{ab} ) {\bf X}_b  =  0\,,
\label{eq:stressg}
\ee
using  the two-dimensional geometrical fact ${\cal R}_{ab} = (1/2) {\cal R} g_{ab}$,
see (\ref{eq:r2d}). In fact, one can read it off simply from  the calculation of the Euler-Lagrange derivative of the Gaussian bending energy (\ref{eq:elg}), where it is shown explicitly  that it vanishes, even before taking the divergence.
The vanishing of the linear stress tensor is a very strong local statement, due to the
fact that the Gaussian bending energy is a topological invariant, yet one may have
expected a non vanishing divergence-free stress tensor.  

The use of Noether's theorem to arrive at the stresses acting on the membranes, even if convincingly convenient, 
may appear too formal. An alternative derivation
of the linear stress tensor for lipid membranes has been offered by Fournier, using the
principle of virtual work, in the height representation \cite{Fournier}. A physically intuitive
approach using balance of forces can be found, for axisymmetric configurations, in the monograph by Evans and Skalak \cite{ES}. A useful comparison with   a microscopic perspective is given by Lomholt and Miao in \cite{LM}, where it is
also addressed the issue of the consequences on the linear stress tensor  of the ambiguity in the Noether current (\ref{eq:ambi}), {\it i.e.} the freedom of adding a gradient. In the Lagrangian formulation, it can be seen as
a canonical transformation. A swift derivation of the stress tensor, using auxiliary variables, has been given by Guven \cite{guvenaux}.

\subsection{Angular stress tensor}

The free energy  is  also invariant under rigid rotations in space. 
Let us consider an infinitesimal rotation of the shape functions,
\be
{\bf W} = \delta_R {\bf X} = {\bf b \times X}\,,
\ee
where ${\bf b}$ is a constant vector. It is convenient first to rewrite the Noether current
${\cal Q}^a$, as defined in
(\ref{eq:emmy}), in terms of the linear stress tensor (\ref{eq:stressf}) and the bending moment  (\ref{eq:fab}), as
\be
{\cal Q}^a (W) =  - \tilde{\bf f}^a
\cdot {\bf W}
- \tilde{\bf f}^{ab}   \cdot {\bf W}_b\,.
\label{eq:emmy2}
\ee
Next, specializing the deformation ${\bf W}$ to a rotation, the Noether current becomes
\be
{\cal Q}^a  (b) = -  {\bf b} \cdot \left[ {\bf X} \times \tilde{\bf f}^a  +
{\bf X}_b \times \tilde{\bf f}^{ab}   \right] = - {\bf b} \cdot \tilde{\bf m}^a\,,
\label{eq:emmy3}
\ee
where, according to Noether's theorem, the conserved  angular stress tensor that follows from rotational 
invariance is given by
\be
\tilde{\bf m}^a = {\bf X} \times \tilde{\bf f}^a  + {\bf X}_b \times \tilde{\bf f}^{ab}\,.
\label{eq:maf}
\ee
This expression clarifies the formal structure in the higher derivative Lagrangian formulation of the angular stress tensor obtained in  \cite{stress02},
recalling that the membrane higher derivative phase space is given by the conjugate pairs
$\{ {\bf X} , \tilde{\bf f}^a ; \xa , \tilde{\bf f}^{ab} \} $.
The first term  in the angular stress tensor is the `orbital' part, the angular stress tensor due to the couple of the linear stress tensor about the origin.
The second term is the pure bending angular stress tensor, due to the couple of the  higher momentum  to the tangent vectors. For a different point of view, this Lagrangian derivation should be compared to the d'Alembertian derivation
 using the principle of virtual work  in the height representation given
by Fournier \cite{Fournier}. 

To see that the angular stress tensor is conserved,  the first variation of the free energy 
is specialized to an  arbitrary connected patch 
$\mathbb{S}_0$ of the surface $\mathbb{S}$, using the same argument given in  \cite{stress02}.  The  invariance of the free energy under rotations 
over the region $\mathbb{S}_0$ is expressed as
\be
0 = \delta_R F [ X] = {\bf b} \cdot \int_{\mathbb{S}_0}  \left[ {\bf X} \times \boldsymbol{\cal E} (\mf)  - \nabla_a \tilde{\bf m}^a  
\right]\,.
\ee
Since both the vector ${\bf b}$ and the patch $\mathbb{S}_0$ are arbitrary, 
this implies the local statement 
\be
 {\bf X} \times \boldsymbol{\cal E} (\mf)  = \nabla_a \tilde{\bf m}^a\,.  
 \label{eq:consa}
 \ee
 Denoting with ${\bf N}$  the torque density acting on the membrane, and using $ \boldsymbol{\cal E} (\mf) = {\bf F}$,
 this equation can be rewritten in the familiar form
 \be
{\bf X} \times {\bf F} = {\bf N}\,.
\ee
Therefore at equilibrium the angular momentum is divergence-free, and the torque density vanishes
\be
{\bf N} = \nabla_a \tilde{\bf m}^a = 0\,.
\ee
Note that the divergence of the angular stress tensor in its form  (\ref{eq:maf})  gives the identity
\be
\xa \times \tilde{\bf f}^a  + \nabla_a \left( {\bf X}_b \times \tilde{\bf f}^{ab}  \right) = 0\,,
\label{eq:ida}
\ee
that can be rewritten in  algebraic form as
\be
\xa \times { \partial \mf \over \partial \xa} + \xab \times {\partial \mf \over \xab } = 0\,,
\ee
to make it the higher derivative field theoretical analogue of the identity $\dot{\bf x} \times {\bf p} = 0$ in the classical mechanics of a point-like object, and the identity follows from the homogeneity of the energy density function.

To illustrate the formalism, and considering that generally the angular momentum tensor is not
taken into proper consideration in the usual treatment of higher-derivative field theories, let us
look at  specific energy densities.  For a soap film the angular stress tensor contains only the orbital part, due to the couple of the linear stress tensor
 about the origin. When the energy density depends on the curvature, there is a non-vanishing
 bending contribution. 
 For the bending energy density ({\ref{eq:bending}), the  angular stress tensor is given by \cite{stress02,Fournier}
 \be
\tilde{\bf m}^a_{(B)} =   {\bf X} \times \tilde{\bf f}^a_{(B)} - 2 \kappa \sqrt{g}\, K g^{ab}  {\bf X}_b \times {\bf n}\,,
 \ee
where the linear stress tensor is given by (\ref{eq:fben}). The  bending part is tangential to the membrane,
and it depends on the mean curvature.

For the mean curvature energy density (\ref{eq:mean}), we have
\be
\tilde{\bf m}^a_{(M)} =   {\bf X} \times \tilde{\bf f}^a_{(M)} - \beta \sqrt{g}\,  g^{ab}  {\bf X}_b \times {\bf n}\,,
 \ee 
 where the linear stress tensor is given by (\ref{eq:fmean2}). 
Note that the second term is automatically divergence-free.

The Gaussian bending energy is a topological invariant, and  its shape equation vanishes identically. 
The linear stress tensor vanishes as well, but the angular stress
tensor,  however, does not \cite{stress04,Fournier}. 
It is given by the  bending angular stress tensor,
\be
\tilde{\bf m}_{(G)}^a = 2 \overline{\kappa}  \sqrt{g} (K g^{ab} - K^{ab} ) {\bf n} \times {\bf X}_b\,.
 \ee
 In the presence of a boundary, or of an edge, in general the Gaussian angular stress tensor will contribute.
 Although not vanishing, the Gaussian bending energy produces no torque density, since it is automatically divergence-free,
 {\it i.e.} it satisfies the identity (\ref{eq:ida}), 
 \ba
\nabla_a\tilde{\bf m}_{(G)}^a  &=& 2 \overline{\kappa} \sqrt{g} \left[  (\nabla^b K - \nabla_aK^{ab} ) {\bf n} \times {\bf X}_b
+  (K g^{ab} - K^{ab} ) K_a{}^c {\bf X}_c \times {\bf X}_b \right. \nn\\
 &-& \left. (K g^{ab} - K^{ab} ) K_{ab}  {\bf n} \times {\bf n}  \right] = 0\,,
\ea
using the contracted Codazzi-Mainardi equation (\ref{eq:cm2}) in the first term.
For the angular stress
tensor, the Gaussian bending contribution is what a field theorist would call 
pure gauge. 

In order to gain insight on the physical meaning of the angular stress tensor, the case of an isolated
membrane is not adequate, and one should consider more complex mechanical problems. 
For a through discussion of the torque balance in the presence of boundaries
and inclusions, and various examples, the reader is again referred to   the review  by Deserno \cite{Deserno15}, and in particular to \cite{MDG,mullerth} for a concrete analysis of the torques exerted by membrane inclusions.

\section{Applications: heterogenous membranes, external forces} 

In this section we extend our considerations to energy densities more general than the energy density 
(\ref{eq:fx1}), with a dependence only on the derivatives of the shape functions, that in the classical mechanics analogy can be  thought of as a `kinetic energy', and consider the general case (\ref{eq:general}), allowing for an explicit
dependance on the shape functions and on the surface local coordinates,
 with an eye towards applications. 
We consider heterogenous membranes, and the addition of local degrees of freedom on the membrane.
We also consider briefly the addition of external forces, described by a `potential energy'.
The two generalizations are kept separate in the presentation, for the sake of clarity.
Our purpose is only  to illustrate with simple examples 
both the convenience and the range of a covariant Lagrangian formulation.

\subsection{Heterogenous membranes}

An heterogenous or anisotropic  membrane can be modeled by position dependent 
bending moduli, and surface tension. The energy density takes the form
\be
\mf = \mf (\xa, \xab, \xi^a )\,.
\ee 
This extension of the Lagrangian formulation  is analogous to the case of an explicitly time dependent Lagrangian in classical mechanics, and it  can be treated accordingly.  
 Because of the breaking of reparametrization invariance, the equilibrium condition acquires a tangential component,
so that the force density is no longer normal to the surface. On the other hand, 
the linear and angular momentum are left 
unchanged in their  form by this generalization. Having allowed for arbitrary, covariant variations from the
very beginning allows to handle this possible description of the  inhomogeneity of the membrane in a straightforward way.
Various investigations have emphasized the importance of this generalization, especially in  
elasticity theory \cite{RossoV,SteigBRB,Agrawal}.

Consider the simple example of a position dependent surface tension $\sigma (\xi)$ for a soap film, due to a 
concentration variation. The energy density
takes the form
$\mf_{(S)} = \sigma (\xi ) \sqrt{g} $. 
Its force density is given by its Euler-Lagrange derivative 
\ba
\boldsymbol{\cal E} (\mf_{(S)}) &=&  - \nabla_a \left({ \partial \mf \over \partial {\bf X}_a } \right) 
= - \nabla_a \left( \sigma (\xi ) \sqrt{g} g^{ab} {\bf X}_b \right) \nn \\ 
&=&  \sigma \sqrt{g} K {\bf n}
- \sqrt{g}  g^{ab}  ( \partial_a \sigma ) {\bf X}_b\,. 
\ea
There is a normal component of the force density, identical to the homogenous case, and an additional tangential
component, due to a gradient in the concentration, that can be interpreted as a Marangoni stress \cite{powers}. 
Note that the linear stress tensor is left unchanged, and it is given by (\ref{eq:fsp}), but with a position
dependent $\sigma (\xi )$, just like a point-like particle with variable mass in classical mechanics.

Let us consider next the case of a 
non-uniform spontaneous curvature with energy density \cite{Bitbol,salac}
\be
 \mf_{(M)} = \beta (\xi ) \sqrt{g}  K\,. 
\ee
It is more instructive to focus first on the linear stress tensor, thinking of its role in problems of adhesion and of contact
in general.
In this case,  the linear stress tensor is given by (\ref{eq:fmean1}), unchanged in its form with respect to the
homogenous case,
\be
\tilde{\bf f}^a_{(M)} =   \sqrt{g} \beta (\xi ) \, \left( 2K^{ab} -   K g^{ab} \right) {\bf X}_b  - \nabla_b \left[ \beta (\xi^a) \sqrt{g} g^{ab} {\bf n} \right]\,.
\ee
Now, however, it is no longer purely tangential as in (\ref{eq:fmean2}), and it acquires a normal component proportional to the gradient of the spontaneous curvature
\be
\tilde{\bf f}^a_{(M)} =   \sqrt{g} \left[ \beta \, \left( K^{ab} -   K g^{ab} \right) {\bf X}_b  -  g^{ab} (\partial_b \beta) 
{\bf n}  \right]\,.
\ee
The force density is given by its divergence as
\be
\nabla_a \tilde{\bf f}^a_{(M)} =   \sqrt{g} \left[  \beta {\cal R} - (\nabla^2 \beta )\right]  {\bf n} -
\sqrt{g} (\partial_a \beta) K^{ab} {\bf X}_b\,. 
\ee
Besides the expected tangential component of the force density of the Marangoni type, also the normal component is modified
by the heterogeneity.

The same pattern emerges for a position dependent bending rigidity $\kappa (\xi )$ in the 
bending energy density  
\be
\mf_{(B)} = \kappa (\xi ) \sqrt{g} \, K^2\,.
\ee
The linear stress tensor is
\ba
 \tilde{\bf f}^a_{(B )} &=&   \sqrt{g} \kappa (\xi) \left( 4 K^{ab} - K g^{ab} \right) {\bf X}_b - 2 \nabla_b \left[ \kappa (\xi) \sqrt{g}
 g^{ab} K {\bf n} \right]
\nn \\
 &=& 
\sqrt{g} \kappa (\xi) \left[ \left( 2 K^{ab} - K g^{ab} \right) {\bf X}_b - 2 g^{ab} (\nabla_b K)  {\bf n} \right]
- 2 \sqrt{g} g^{ab} K (\partial_b \kappa ) {\bf n}. 
\ea
It is given by the stress tensor of the homogenous case with the addition of a normal component
proportional to the gradient of the bending rigidity. The force density given by the divergence of this stress tensor is easily calculated, but it is not particularly illuminating.

The case of position dependent Gaussian  bending rigidity is of interest because the inhomogeneity simply destroys 
the topological invariance of the Gaussian bending free energy. Consider the energy density 
\be
\mf_{(G)} = \overline{\kappa} (\xi ) \sqrt{g} \, {\cal R}\,.
\ee
The linear stress tensor for the homogenous case vanishes, see ({\ref{eq:stressg}),  but now it takes the form 
\be
\tilde{\bf f}^a_{(G)} =   2 \sqrt{g} \, \left( K^{ab} -   K g^{ab} \right)  (\partial_b \overline\kappa )\,
{\bf n}\,,  
\ee
with only a normal component. The force density is also non-vanishing
\be
\nabla_a \tilde{\bf f}^a_{(G)}  =   2 \sqrt{g} \, \left( K^{ab} -   K g^{ab} \right)  (\nabla_a \nabla_b \overline\kappa ) \, {\bf n}
+ \sqrt{g} g^{ab} {\cal R}  (\partial_a \overline\kappa )  {\bf X}_b\,,
\ee
where we have used the Gauss-Codazzi-Mainardi equation (\ref{eq:g2}), and the two-dimensional fact ${\cal R}_{ab} = (1/2) {\cal R} g_{ab}$ to obtain the tangential component of the Marangoni type; the derivation is immediate, but not obvious.

If to let the phenomenological parameters become position dependent is straightforward, 
so it is to consider internal fields living on the membrane. This is of interest in the general topic
of active membranes, where one is interested in additional degrees of freedom localized on the
surface representing the membrane, see {\it e.g.} the recent work on the mechanics of active
membranes by Salbreux and J\"{u}licher \cite{julicher}. 

We consider a single example of
a different modeling of the heterogeneity,  given  in terms of a phase field $\phi = \phi (\xi^a)$,
describing a variation in the lipid concentration. A simple description  is a Ginzburg-Landau type scalar
field  with a reparametrization covariant  energy density \cite{stress04,Napolieq, Desernoch}
\be
\mf_{(\phi)} = \sqrt{g} \left[ {\lambda \over 2} g^{ab} \nabla_a \phi \nabla_b \phi + V (\phi)  + \beta_{(\phi)} \phi K \right]\,,
\ee
where $\lambda, \beta_{(\phi)}$ are coupling constants, and the last term represents a coupling of the 
phase field with the mean curvature.  With the Lagrangian  formulation, we obtain in an efficient way
the linear stress tensor for this model, and the surface force density it produces. 
We assume that the scalar field, satisfies its own equilibrium condition, with appropriate boundary conditions,  given by
\be
\lambda \nabla^2 \phi - V'(\phi) = \beta_{(\phi)} K\,.
\label{eq:phieq}
\ee
 We calculate the partial derivatives
\ba
\ {\partial \mf_{(\phi)} \over \partial {\bf X}_a } &=& \sqrt{g} \left[ - \lambda \nabla^a \phi \nabla^b \phi +
{\lambda \over 2} (\nabla \phi )^2 + V (\phi ) g^{ab} + \beta_{(\phi)} \phi  (K g^{ab} - 2 K^{ab} )  \right] {\bf X}_b\,,  \nn
\\
{\partial \mf_{(\phi)} \over \partial {\bf X}_{ab} } &=& - \beta_{(\phi)} \phi \sqrt{g} g^{ab} {\bf n}\,. \nn
\ea
The linear stress tensor due to the scalar  is obtained immediately  \cite{CG04}
\ba
\tilde{\bf f}^a_{(\phi)} = \sqrt{g} \left[ \lambda \nabla^a \phi \nabla^b \phi - {\lambda \over 2} (\nabla \phi )^2  - g^{ab} V (\phi) - \beta_{(\phi)} \phi  (K g^{ab} -  K^{ab} ) \right] {\bf X}_b\,,  \nn \\
\qquad \qquad - \beta_{(\phi)} \sqrt{g} g^{ab}  (\nabla_b \phi ) {\bf n}\,. 
\ea
Its divergence gives the Euler-Lagrange derivative, or force density, using the equilibrium condition
for the scalar field (\ref{eq:phieq}), 
\be
 \boldsymbol{\cal E} (\mf_{(\phi)}) = \sqrt{g} \left[ - \lambda K^{ab} \nabla_a \phi \nabla_b \phi + {\lambda \over 2} K (\nabla \phi )^2 
+ K V (\phi) + \beta_{(\phi)} ( \phi {\cal R} - \nabla^2 \phi ) \right] \, {\bf n}\,.
\ee 
The force density is normal to the surface because of reparametrization invariance.

\subsection{External forces and constraints}

In the presence of external forces,  the energy density  can possess an explicit dependence on the
shape functions. In the classical mechanics analogy it  can be considered as a
potential energy,  and the energy density can be written as
\be
\mf =  \mf ({\bf X},  \xa, \xab )\,.
\ee
In the Lagrangian formulation,  the variational differential (\ref{eq:fo1}) is modified to include 
the extra dependence to 
\be
\delta  \mf =  {\partial \mf \over \partial {\bf X}} \cdot {\bf W} + {\partial \mf \over \partial {\bf X}_a } \cdot {\bf W}_a + {\partial \mf \over \partial {\bf X}_{ab } } \cdot {\bf W}_{ab}\,.  
\label{eq:fog}
\ee
This is the fundamental relationship from which one could say that the rest follows, according to the rules of the calculus 
of variations. The first variation of the free energy is left unchanged, and it is the same as (\ref{eq:varf1}), 
but the Euler-Lagrange derivative 
takes now the form
\be
 \boldsymbol{\cal E} (\mf) =  {\partial \mf \over \partial {\bf X} } 
 - \nabla_a \left({ \partial \mf \over \partial {\bf X}_a } \right) 
+  \nabla_a \nabla_b \left( {\partial \mf \over \partial {\bf X}_{ab} } \right)\,.
\label{eq:elo}
\ee
On the other hand, the Noether current (\ref{eq:emmy}) is clearly left  unchanged in its form. 
However, in general, the linear and the angular momentum stress tensors will be modified, since the energy density
representing the external force will depend in a non trivial way on the geometry of the surface.

The equilibrium condition is now given by the vanishing of the Euler-Lagrange derivative (\ref{eq:elo}).
Using the linear stress tensor, it can be written alternatively  as 
\be
\nabla_a \tilde{\bf f}^a = - {\partial \mf \over \partial {\bf X}}\,.
\ee
The linear stress tensor is no longer conserved, of course, and moreover one needs to keep in mind the choice for 
the sign of the linear stress tensor.

A simple, and physically relevant, example is an isotropic osmotic pressure acting on a closed membrane.
This is described  by  a free energy proportional to the volume enclosed,
\be
V = \int_{int \mathbb{S}} d^3 x = {1 \over 3} \int_{\mathbb{S}} \sqrt{g} \, {\bf n} \cdot {\bf X}\,,
\ee
with energy density
\be
\mf_{(V)} = 
{1 \over 3} \sqrt{g} \, {\bf n} \cdot {\bf X}\,, 
\ee
that can be added to the membrane energy density as a constraint, enforcing constant volume, as
\be
\mf \to \mf_{(c)} = \mf - P \mf_{(V)}\,,
\label{eq:fc}
\ee
where $P$ denotes the pressure. If this addition is seen as a constraint that enforces 
constant volume because of the low permeability of the membrane, then
the pressure $P$ can be seen as a Lagrange multiplier. 
To derive the contribution of the external pressure to the equilibrium condition, we  calculate the partial derivatives
\ba
{\partial \mf_{(V)} \over \partial {\bf X}_a } 
&=& {1 \over 3} \sqrt{g} \left[ ( {\bf n} \cdot {\bf X} ) {\bf X}^a -  ( {\bf X}^a \cdot {\bf X} ) {\bf n} \, \right] \label{eq:vf} \\
{\partial \mf_{(V)} \over \partial {\bf X} } &=& {1 \over 3} \sqrt{g} \; {\bf n}\,,
\ea
and therefore the contribution to the force density due to the external pressure is
\be
\boldsymbol{\cal E} (\mf_{(V)}) = {1 \over 3}  \sqrt{g} \,  {\bf n} - {1 \over 3} \sqrt{g} \nabla_a \left[ ( {\bf n} \cdot {\bf X} ) {\bf X}^a -  ( {\bf X}^a \cdot {\bf X} ) {\bf n} \right] =   \sqrt{g} \, {\bf n}\,, 
\ee
using the Gauss-Weingarten equations 
(\ref{eq:gw1}) and (\ref{eq:gw2}), and the identity ${\bf X}^a \cdot {\bf X}_a = 2 $ in the calculation of the divergence. 

When the constraint of constant volume is added, for example, to the soap film free energy as in (\ref{eq:fc}),
the shape equation becomes
\be
\boldsymbol{\cal E} (\mf_{(c)}) = \boldsymbol{\cal E} (\mf_{(s)}) - P \boldsymbol{\cal E} (\mf_{(V)}) 
= \sqrt{g} \left[ \sigma K - P  \right] \, {\bf n} = 0\,,
\ee   
and one recognizes the familiar Young-Laplace equation for soap bubbles, $\sigma K = P$.
If the constraint of constant volume is imposed on the
 Canham-Helfrich model as in (\ref{eq:fc}), the shape equation  is modified to
\be
\boldsymbol{\cal E} (\mf_{(c)}) = \boldsymbol{\cal E} (\mf_{(CH)}) - P \boldsymbol{\cal E} (\mf_{(V)}) 
= \boldsymbol{\cal E} (\mf_{(CH)}) - P \sqrt{g} \, {\bf n} = 0\,,
\ee   
where the Euler-Lagrange derivative for the Canham-Helfrich model is given by the left hand side of (\ref{eq:chs}), and the 
pressure adds a source term.
It should be mentioned that this equilibrium condition in the presence of an
external pressure is the original `shape equation' for lipid vesicles introuced by Ou-Yang and Helfrich  \cite{Hel87,Hel89}. 

Note that the partial derivative  (\ref{eq:vf}) gives directly, no derivation needed, a  linear stress tensor for the enclosed volume
\be
\tilde{\bf f}^a_{(V)} =   {1 \over 3} \sqrt{g} g^{ab}  \left[ ( {\bf X}_b \cdot {\bf X} ) {\bf n} - ( {\bf n} \cdot {\bf X} ) {\bf X}_b  
\right]\,. 
\ee
This linear stress tensor, up to a numerical factor, has been 
used by Guven in \cite{guvlaplace} to  give an interesting interpretation of the Laplace pressure in fluid vesicles, see also M\"{u}ller in \cite{mullerth}.

As  a second  example  of an external force acting on a lipid membrane, we  consider a spatially varying external magnetic field ${\bf B} ({\bf X})$, to illustrate the formulation.  The lipid membrane is considered as a diamagnetic material, and the effect of the
magnetic field on the lipid constituents is described by an alignment energy density \cite{Helfm,salac16,Iwa}
\be
\mf_{(m)} = - \alpha_{(m)}\,\sqrt{g} \; [{\bf n} \cdot {\bf B} ({\bf X} ) ]^2\,,
\ee
where for the sake of brevity we use a phenomenological constant  $\alpha_{(m)} = \Delta \chi d / 2 \mu_m$, where $d$ is the membrane thickness, $\mu_m$ the membrane magnetic permeability, and   $\Delta \chi = \chi_\parallel - \chi_\perp$ the difference between the magnetic susceptibilities in the parallel and perpendicular direction of the lipid molecules.
We follow the recent study by Salac in \cite{salac16}.

Using the Lagrangian formulation, we derive readily the linear stress tensor and the force density 
produced by this magnetic energy density.
We have that 
\ba
{\partial \mf_{(m)} \over \partial {\bf X}_a } &=&  - \alpha_{(m)}  \,\sqrt{g} ( {\bf n} \cdot {\bf B} ) g^{ab}
[ ( {\bf n} \cdot {\bf B} ) {\bf X}_b - 2 ( {\bf X}_b \cdot {\bf B} ){\bf n} ]\,,
\label{eq:mp1}
\nn \\
{\partial \mf_{(m)} \over \partial {\bf X}} &=&  - 2 \alpha_{(m)}  \,\sqrt{g} ( {\bf n} \cdot {\bf B} ) n_\mu \vec{\bf \nabla} B^\mu\,.
\nn \ea
The linear stress tensor of the magnetic energy density is obtained immediately  from the general expression  (\ref{eq:stressf}) and  the partial derivative (\ref{eq:mp1}) as
\be
\tilde{\bf f}^a_{(m)}  =  \alpha_{(m)}\,\sqrt{g} g^{ab} ( {\bf n} \cdot {\bf B} )
[ ( {\bf n} \cdot {\bf B} ) {\bf X}_b - 2 ( {\bf X}_b \cdot {\bf B} ){\bf n} ]\,.
\ee
The Euler-Lagrange derivative of the alignment energy density gives the force density
in the form
\be
\boldsymbol{\cal E} (\mf_{(m)}) = \nabla_a  \tilde{\bf f}^a_{(m)}  - 2 \alpha_{(m)}  \,\sqrt{g} ( {\bf n} \cdot {\bf B} ) n_\mu \vec{\bf \nabla} B^\mu\,,
\ee
where we use $\vec{\bf \nabla}$ to denote the spatial gradient.

Let us specialize to a spatially constant magnetic field for the sake of simplicity. In this case, the force density is given by 
\be
\nabla_a \tilde{\bf f}^a_{(m)}  = \alpha_{(m)} \,\sqrt{g} \left[ K   ( {\bf n} \cdot {\bf B} )^2  -  2 K^{ab}
( {\bf X}_a \cdot {\bf B} ) ({\bf X}_b \cdot {\bf B} ) \right] {\bf n}\,.
\ee
In this specialization, reparametrization invariance is restored, and the force density is normal to the surface.
We recover the expression derived by Salac in \cite{salac16}, that uses in the derivation the Napoli-Vergori method \cite{Napolieq}. 

\subsection{Remark on lipid domains}

A different sort of inhomogeneity in lipid membranes is given by 
domains, characterized by a sharp phase separation of the lipid constituents.
Consider a binary membrane, made of two different lipids, that we label with $\alpha,\beta$, 
 described by a surface $\mathbb{S}$ given by the union of
two regions $\mathbb{S}_{(\alpha)}$ for the first lipid, and $\mathbb{S}_{(\beta)}$ for the second.
The two regions share a common simply connected boundary, described by a curve ${\cal C}$. 
Our purpose here is only to remark on the role played by the linear stress tensor and the
bending momentum  in the determination of the appropriate boundary conditions on ${\cal C}$.
The free energy for the binary membrane can be written as the sum
\be
F [X] = \int_{\mathbb{S}_{(\alpha)}} \mf_{(\alpha)} +  \int_{\mathbb{S}_{(\beta)}} \mf_{(\beta)}\,,
\ee
where each phase is described by some energy densities, $\mf_{(\alpha)}$ and $\mf_{(\beta)}$.
The first variation of the free energy, using (\ref{eq:varf1}), is
\be
\delta F =   \int_{\mathbb{S}_{(\alpha)}}\,  \left[  \boldsymbol{\cal E} (\mf_{(\alpha)})  \cdot {\bf W}  
+ \nabla_a {\cal Q}_{(\alpha)}^a (W) \right]  +
\int_{\mathbb{S}_{(\beta)}}\,  \left[  \boldsymbol{\cal E} (\mf_{(\beta)})  \cdot {\bf W}  
+ \nabla_a {\cal Q}_{(\beta)}^a (W) \right]\,. 
\ee
Let us focus on the boundary contribution to the variation, to see how the appropriate boundary
conditions emerge in the Lagrangian formulation. 
Using the expression (\ref{eq:emmy2}) for the Noether current in terms of the linear stress tensor and and the
bending momentum, the boundary part of the  first variation becomes 
\ba
\delta F [X] &=& \int_{\mathbb{S}_{(\alpha)}} \nabla_a {\cal Q}_{(\alpha)}^a (W)  +
\int_{\mathbb{S}_{(\beta)}} \nabla_a {\cal Q}_{(\beta)}^a (W) 
\nn \\
&=& \oint_{\cal C} ds \;  l_a \left[ \left( {\bf f}^a_{(\beta)}  -  {\bf f}^a_{(\alpha)}  \right)  
\cdot {\bf W}
+ \left(  {\bf f}^{ab}_{(\beta)} -  {\bf f}^{ab}_{(\alpha)}  \right) \cdot {\bf W}_b \right]\,.
\ea
where $l^a$ denotes the unit normal to the curve pointing from  $\mathbb{S}_{(\alpha)}$ to $\mathbb{S}_{(\beta)}$, and
 we omit the tilde in the linear stress tensor and  in the bending moment because the
line integral is over arc-length. In the classical mechanical analogy, this expression can be interpreted  formally as 
the familiar conservation of the linear momentum and of the higher linear momentum at the boundary.
For the  use of the consequences of this expression in the study of the adhesion of two lipid membrane, see Deserno, M\"{u}ller, and Guven 
in \cite{contact}, and the review by Deserno  \cite{Deserno15}. 

\section{Second variation}

The second variation of the free energy functional about equilibrium is the first step in the  study of the issue of the local stability of lipid  membranes.  
The stability analysis for lipid membrane models has been the subject of various investigations. 
For the Canham-Helfrich model, the study of the second variation was considered  using differential geometric 
tools by
 Ou-Yang and Helfrich \cite{Hel87,Hel89},  
with applications to spherical and cylindrical shapes, see also  \cite{Pet2,Bukman}. 
The general structure of the second variation for arbitrary solutions has also been 
explored, using a covariant geometric approach   \cite{Ou,CGS03,CG04,Y2L,Waladi}. 
In continuum mechanics, it has been the subject of rigorous analysis, in particular in the work
by Steigmann \cite{Steigeq}.
 A one-dimensional reduction of the Canham-Helfrich model for the case of adhering vesicles has provided
useful insight \cite{Rossosec}. The second variation appears also in the standard perturbative calculation of the renormalization of the bending rigidities due to thermal fluctuations, to produce the effective potential \cite{Jerusalem}.

In this section, the Lagrangian formulation is put to use to obtain quite efficiently 
a general expression for the second variation, for an energy density that depends
at most on second derivatives of the shape functions, of the form (\ref{eq:fx1}), that comes tantalizingly close
to its analogue in classical mechanics. 
A previous investigation, reported in \cite{CG04}, uses a covariant expansion of the geometric
scalars that appear in the   energy density to second order in the deformations, and provides a useful benchmark.
The Lagrangian formulation allows to better appreciate the structure of the second variation, and it 
makes possible a  generalization to more general settings, as heterogenous membranes,
or mechanical systems more complex than an isolated closed membrane. Constraints like constant
area and constant volume can be easily included, as problems involving adhesion and contact, or membranes
with open or loaded edges \cite{Edge,Tuedge}. 
In all these instances, an expression for the second variation of the free energy is the first necessary step
towards any study of local stability and the issue of shape transformations.

We assume that the equilibrium condition is satisfied, $\delta F [ X] = 0$.  The strategy we use is to derive directly the second variation of the energy density. An alternative, and equivalent strategy would be to focus on the
linearization of the Euler-Lagrange derivative, but this would require integrations by parts that we wish to avoid, if possible,
thinking of applications to more complex mechanical problems than an isolated closed membrane. Firstly,
we consider the first covariant variational derivative of the energy density, using  (\ref{eq:fo1}),  along a deformation
vector $W$. Secondly we perform  a second variation along a different direction $Z$, for consistency, {\it i.e.} $\delta_Z \delta_W \mf $,   letting them coincide afterwards.
In order not to clutter the notation, we write the 
covariant second variational derivative of the energy density in an implicit form as
\be
\delta^2 \mf  =  \delta \left[ 
{\partial  \mf \over \partial {\bf X}_a } \cdot {\bf W}_a
+ {\partial  \mf \over \partial {\bf X}_{ab}} \cdot {\bf W}_{ab} \right]\,.
\label{eq:sd1}
\ee
At this point, it is important to observe  that the affine connection appears implicitly in the second derivative of the variation,
and that this is due to the higher-derivative nature of the theory under consideration.
Taking this fact into account, in the coincidence limit  $Z \to W$,  one obtains the explicit expression for the
second variation of the energy density
\ba
 \delta^2 \mf = \left( {\partial^2  \mf \over \partial X^\nu_c \partial X^\mu_a } \right)   W^\mu_a W^\nu_c + 
  \left( {\partial^2  \mf \over \partial X^\nu_c \partial X^\mu_{ab}  } \right) W^\mu_{ab} W^\nu_c 
  +  \left( {\partial^2  \mf \over \partial X^\nu_{cd } \partial X^\mu_a } \right)  W^\mu_a W^\nu_{cd}   
\nn\\
 +  \left({\partial^2  \mf \over \partial X^\nu_{cd} \partial X^\mu_{ab} } \right) W^\mu_{ab}  W^\nu_{cd} 
- \left( {\partial \mf \over \partial  X^\mu_{ab } } \right)  (\delta \Gamma^c{}_{ab} ) W^\mu_c\,.
\label{eq:sd3}
\ea
Here spatial greek indices are necessary. 
 All possible Hessians of the energy density appear, as one would have expected from the analogy with a 
 Lagrangian formulation of a system with a finite number of degrees of freedom \cite{Rund,GF,GH}. 
 What spoils the nice purely Hessian structure is on one hand the fact that the mixed partial derivatives do not
commute. This is a consequence of the fact that the basis $\{ (\partial/ \partial {\bf X}_a) , (\partial/ \partial {\bf X}_{ab} ) \} $
for the phase space tangent space is not a coordinate basis. A second spoiler
is the variation of the affine connection in the last term.
Since the affine connection is not a function on the Lagrangian 
phase space, the variational differential does not apply, and its variation has to be 
considered separately. This is the price to pay for insisting on general covariance in the election of the phase space field
variables. Of course, this issue
does not arise for an intrinsic model, like a soap film, and it can be sidestepped for a model linear in
second derivatives, but not for the bending energy density. 
However, the price is not too steep. Under a deformation of the shape functions, 
a short calculation
gives that  the variation of the affine connection is 
\ba
\delta \Gamma^c{}_{ab} &=&
\delta  \left( g^{cd} {\bf X}_d \cdot \partial_a {\bf X}_b \right)   \nn \\
 &=& {\bf X}^c \cdot \partial_a {\bf W}_{b}  - ( {\bf X}^c \cdot {\bf W}^d ) {\bf X}_d \cdot \partial_a {\bf X}_b 
+ {\bf W}^c \cdot \partial_a {\bf X}_b 
- ({\bf X}^d \cdot {\bf W}^c)   {\bf X}_d \cdot \partial_a {\bf X}_b 
\nn \\
 &=& {\bf X}^c \cdot {\bf W}_{ab} + ({\bf n} \cdot {\bf W}^c )  {\bf n}  \cdot \partial_a {\bf X}_b
= {\bf X}^c \cdot {\bf W}_{ab} -   K_{ab} \, {\bf n} \cdot {\bf W}^c\,,
\label{eq:ac}
\ea
where we have used the variation of the
covariant metric 
$
\delta g^{ab} = - 2 {\bf X}^{(a} \cdot {\bf W}^{b)}
$ in the first line, and 
the completeness relationship (\ref{eq:comple}) to obtain the third line. 

Thus, we arrive rather quickly to a general expression for the second variation of the free energy as
\ba
\qquad \delta^2 F [ W ] = \int_{\mathbb{S}} \delta^2 \mf 
= \int_{\mathbb{S}} \left[ \left({\partial^2  \mf \over \partial X^\nu_{cd} \partial X^\mu_{ab} } \right) W^\mu_{ab}  W^\nu_{cd}  
+ 
  \left( {\partial^2  \mf \over \partial X^\nu_c \partial X^\mu_{ab}  } \right) W^\mu_{ab} W^\nu_c 
\right. \nn\\
 \left.\qquad  \qquad +  \left( {\partial^2  \mf \over \partial X^\nu_{cd } \partial X^\mu_a } \right)  W^\mu_a W^\nu_{cd}   
+  
\left( {\partial^2  \mf \over \partial X^\nu_c \partial X^\mu_a } \right)   W^\mu_a W^\nu_c 
\right.\nn \\
\qquad \qquad \left. 
- \left( {\partial \mf \over \partial  X^\mu_{ab } } \right) X_\nu^c  \, W^\nu_{ab}  W^\mu_c 
+ \left( {\partial \mf \over \partial  X^\mu_{ab } } \right) K_{ab} n_\nu  \,  W^{\nu\,c}  W^\mu_c 
\right]\,,
\label{eq:svar1}
\ea
where the third line is the contribution due to the variation of the affine connection given by inserting (\ref{eq:ac})
in (\ref{eq:sd3}).
This general expression for the second variation of the free energy about equilibrium  is the central result of this section; it 
gives
 a bilinear form on the space of perturbations about equilibrium, and it appears in a different guise in previous 
 works on the subject. One benefit of this Lagrangian derivation is that  it makes apparent
 the dependence on the energy density. 
 For any given energy density, the evaluation of the second variation  is reduced to the calculation
 of the Hessians, and it can be easily specialized to specific classes of configurations, like spherical,
 cylindrical, or axisymmetric. Moreover,  the derivation avoids integration by parts, and so it does not
 require to make any assumption about boundary conditions, as it stands. This means that it is
 applicable also in the presence of edges or in problems of adhesion, see \cite{Rossosec,rossoproc}.
 
 Positivity of the bilinear form, or $ \delta^2 F [W] \geq 0 $ about equilibrium  indicates stability, but any
study of the stability issue requires a careful understanding of all the allowed perturbations,
where the seemingly innocent word `allowed' hides  a plethora of 
mathematical subtleties \cite{GF,GH}. Moreover, in general 
the issue of the election of a suitable basis of test functions for the perturbations depends on the
concrete problem at hand. Where the availability of a general expression for the second variation should be of assistance
is when comparing different modes of possible perturbations about equilibrium, and a quantitative 
assessment of which one is energetically favoured.

The top Hessian is given by the second partial derivative of the energy density with respect to
 the second derivatives. It can be used to provide a Legendre-Hadamard necessary condition for stability
 \cite{GH}
  \be
\left({\partial^2  \mf \over \partial X^\nu_{cd} \partial X^\mu_{ab} } \right) W^\mu_{ab}  W^\nu_{cd}  \geq 0\,,
\label{eq:LH}
\ee
that to be implemented requires a definite choice of a basis for the perturbations.
An equivalent condition has been derived by Steigmann using elasticity theory in  \cite{Steigeq}.

In the following, our aim is to consider specific examples, both to understand and to  check the validity of the general expression (\ref{eq:svar1}) for specific
energy densities, using the results of \cite{CG04} as a guidance. 

To consider a concrete example, for a soap film, the energy density (\ref{eq:soap}) is proportional to the area,
and therefore there is only one Hessian,
\be
{\partial^2 \mf_{(S)} \over \partial X^\mu_a \partial X^\nu_b } =  \sigma
\sqrt{g} \left( g^{ab} n_\mu n_\nu + X^{ab}_{\mu\nu} \right)\,.  
\label{eq:hss}
\ee
 The Hessian has  a normal and a tangential part. 
 The tangential part is proportional to the tangential bivector
 \be
  X^{\mu\nu}_{ab} =  X^\mu_a X^\nu_b - X^\mu_b X^\nu_a\,,  
 \label{eq:biv}
 \ee  
 with all indices raised and lowered with the contravariant metric $g^{ab}$ and the Kronecker delta.
As a six by six matrix, the Hessian (\ref{eq:hss})  has vanishing determinant. This is a direct  consequence of reparametrization 
invariance, and indicates the presence of constraints in the phase space, just like for any system
with a gauge symmetry.

The second variation of the soap film energy density, from (\ref{eq:svar1}) and (\ref{eq:hss}),  is then  
\be
\delta^2 F_{(S)}  [ W ]  = \int_{\mathbb{S}}  \sqrt{g}  \sigma \left( g^{ab} n_\mu n_\nu + X^{ab}_{\mu\nu} \right)  
W^\mu_a W^\nu_b\,,  
\label{eq:ss2}
\ee
and this identifies the general form of the bilinear form for minimal surfaces.
In the case of an isolated closed minimal surface, one can consider only a normal deformation ${\bf W} = \phi \, {\bf n}$,
since the tangential components of the deformation contribute only a boundary term \cite{FT,CG04}.
The second variation (\ref{eq:ss2})  takes the form 
\ba
\delta^2 F_{(S)}  [ W ]&=& \int_{\mathbb{S}}  \sqrt{g} \sigma \left[ \nabla^a \phi \nabla_a  \phi + \left(K^2 - K^{ab} K_{ab} \right) \phi^2  \right] \nn
\\ 
&=&
 \int_{\mathbb{S}} \sqrt{g} \sigma \left[ \nabla^a \phi \nabla_a  \phi + {\cal R} \, \phi^2  \right]\,,
\ea
where we have used the Gauss theorem (\ref{eq:g3}) to obtain the second line. 
This is the starting step in the analysis of the local stability of tension dominated systems. In the mathematics
of minimal surfaces, it is used  in the study of the small deformations of the surfaces, 
and in  the calculation of the indices of minimal surfaces,
where by index it is meant the 
characterization of the degree of degeneracy and negative definiteness of the Hessian \cite{FT,minimal}.

Turning to the elastic bending energy density (\ref{eq:bending}), we have four Hessians to calculate. The top Hessian is  
\be
{\partial^2 \mf_{(B)} \over \partial X^\mu_{ab} \partial X^\nu_{cd} }
= 2 \kappa \sqrt{g} g^{ab} g^{cd} n_\mu n_\nu\,.
\ee
Also in this case, as a nine by nine matrix, the top Hessian has vanishing determinant, because of   reparametrization 
invariance, and this fact implies the presence of constraints on the membrane phase space.
In the context of lipid membranes,
this issue is addressed in the canonical Hamiltonian formulation of \cite{Ham1}.
The Legendre-Hadamard necessary condition (\ref{eq:LH}) implies simply positivity of the bending
rigidity $\kappa$.

The other three Hessians appearing in (\ref{eq:svar1}) require more work, 
\ba
 {\partial^2 \mf_{(B)} \over \partial X^\nu_c \partial X^\mu_{ab} }
&=& 2 \kappa \sqrt{g} \left[ \left( -  K g^{cd} g^{ab} 
+ 4 K g^{c(a} g^{b)d} + 2 K^{ab} g^{cd}  \right)  X_{\nu\,d} \,  n_\mu +   K g^{ab} n_\nu X_{\mu}^a \right]\,,
\label{eq:a1} \nn
\\
 {\partial^2 \mf_{(B)} \over \partial X^\nu_{cd} \partial X^\mu_a }
&=& 2 \kappa \sqrt{g} \left[ \left( -  K g^{cd} g^{ab} 
+ 4 K g^{a(c} g^{b)d} + 2 K^{cd} g^{ab}  \right)  X_{\mu\,b} n_\nu  \right]\,,
\label{eq:a2} \nn
\ea
\ba
{\partial^2 \mf_{(B)} \over \partial X^\mu_a \partial X^\nu_b} 
= \kappa \sqrt{g} \left[  K (K g^{ab} - 4 K^{ab} ) n_\mu n_\nu 
+  4 ( K K^{cd} g^{ab} + 2 K^{ac} K^{bd} ) X_{\mu\, c } X_{\nu \, d}  \right. \nn \\
\left. + K^2 X_{\mu\nu}^{ab} 
+ 4 K K_c{}^a X^{bc}_{\mu\nu} - 4 K K_c{}^b X^{ac}_{\mu\nu} 
\right]\,. \nn
\label{eq:a3}
\ea   
In the last line the tangential bivector is given by (\ref{eq:biv}). Note that the Hessians with mixed partial derivatives 
(\ref{eq:a1}), (\ref{eq:a2}) differ, because the basis in the tangent space of the phase space is not  a coordinate
basis.
We also need the extra terms that come from the variation of the affine connection (\ref{eq:ac}) and that appear in
(\ref{eq:svar1})
\ba
- \left( {\partial \mf_{(B)} \over \partial  X^\mu_{ab } } \right) X_\nu^c  \, W^\nu_{ab}  W^\mu_c 
+ \left( {\partial \mf_{(B)} \over \partial  X^\mu_{ab } } \right) K_{ab} n_\nu  \, W^\nu_d  W^\mu_c \nn \\ 
=
2 \kappa \sqrt{g} \left[ K ( {\bf n} \cdot {\bf W}_c )  ({\bf X} ^c \cdot \nabla^2 {\bf W} ) 
- K^2 ({\bf n} \cdot {\bf W}_c ) ({\bf n} \cdot {\bf W}^c ) \right]\,.
\label{eq:acb}
\ea

We arrive  therefore, using the Hessians and (\ref{eq:acb}) in the general expression (\ref{eq:svar1})
at the second variation of the bending energy density, that written in full glory is  
\ba
\delta^2 F_{(B)}  [ W ]  =  2 \int_\mathbb{S}  \sqrt{g}  \kappa \left\{   ( {\bf n} \cdot \nabla^2 {\bf W} )^2
+ 2 (2 K^{ab} - K g^{ab} ) ( {\bf X}_a \cdot {\bf W}_b ) ({\bf n} \cdot \nabla^2 {\bf W} ) \nn \right.  \\
 \left. + 4 K ( {\bf X}^b \cdot {\bf W}^a ) 
 ({\bf n} \cdot  {\bf W}_{ab} ) 
 - {1 \over 2}  K^2 ( {\bf n} \cdot {\bf W}^a ) ({\bf n} \cdot {\bf W}_a ) \right. \nn \\
 \left. - 2 K K^{ab} ( {\bf n} \cdot {\bf W}_a ) ({\bf n} \cdot  {\bf W}_b )
  + 4 K^{ab} K^{cd}( {\bf X}_a \cdot {\bf W}_b ) ({\bf X}_c \cdot {\bf W}_d ) 
\right. \nn \\
\left. + 2 K K^{cd} ( {\bf X}_d \cdot {\bf W}^a ) ({\bf X}_c  \cdot  {\bf W}_a ) 
+ 2 K ({\bf n} \cdot {\bf W}_a ) ({\bf X}^a \cdot \nabla^2 {\bf W} ) \right. \nn \\ 
+ \left. {1 \over 2}  K^2 [ ( {\bf X}^a \cdot {\bf W}_a)^2  - ({\bf X}^a  \cdot  {\bf W}^b ) ({\bf X}_b  \cdot  {\bf W}_a )    ] \nn
\right.
\\
\left.  + 4 K K^{ab} ( {\bf X}_b \cdot {\bf W}_a) ({\bf X}^c  \cdot  {\bf W}_c )  
 - 4 K K^{bc} ( {\bf X}_c \cdot {\bf W}_a) ({\bf X}^a  \cdot  {\bf W}_b )   \right\}\,.
 \ea
As expected, this second variation result is quite complicated. 
 It reproduces  eq. (104) of \cite{CG04}, obtained  by a geometric expansion to 
second order of the bending free energy, whereas here it is obtained from the general expression (\ref{eq:svar1}). 
This can be seen as a non trivial check. 
Which approach to take is clearly  only a question of preference, all one can say is that the work involved is
distributed in a different way. For the specialization to purely normal deformations  see \cite{CGS03}, where,
by judicious integration by parts, it is shown that the tangential deformations appear only in total divergence terms.

To include the spontaneous curvature, let us also consider the second variation of the area difference functional (\ref{eq:mean}). 
In this case the top Hessian vanishes, because of the linearity in the second derivatives of the energy density.
The other Hessians are  given by
\be
{\partial^2 \mf_{(M)} \over \partial X^\nu_{cd} \partial X^\mu_a}
= \beta \sqrt{g} \left[  -  g^{cd} g^{ab} 
+ 2 g^{a(c} g^{d) b} \right] n_\nu X_{\mu\, b}\,,
\ee

\be
{\partial^2 \mf_{(M)} \over  \partial X^\nu_{c} \partial X^\mu_{ab}} 
=  \beta \sqrt{g} \left[ \left( -  g^{ab} g^{cd} 
+ 2 g^{c(a} g^{b) d} \right) n_\mu X_{\nu\, b} + g^{ab} n_\nu X_\mu^c \right]\,,
\ee

\ba
{\partial^2 \mf_{(M)} \over \partial X^\mu_a \partial X^\nu_c} 
= \sqrt{g} \left[   (K g^{ac} - 2K^{ac} ) n_\mu n_\nu 
+ 2 g^{ac} K^{de} X_{\mu\,d} X_{\nu\,e}
\nn \right.\\
\qquad \qquad \left. +  K \, X_{\mu\nu}^{ac} + 2  K_d{}^a X^{cd}_{\mu\nu} - 2 K_d{}^c X^{ad}_{\mu\nu} 
\right]\,,
\ea   
where in the last line the tangential bivector is given by (\ref{eq:biv}). 
Again, note that the Hessians with mixed partial derivatives differ. 
We also need the terms that come from the variation of the affine connection (\ref{eq:ac}),
that in this case take the form 
\ba
 - \left( {\partial \mf_{(M)} \over \partial  X^\mu_{ab } } \right) X_\nu^c  \, W^\nu_{ab}  W^\mu_c 
+ \left( {\partial \mf_{(M)} \over \partial  X^\mu_{ab } } \right) K_{ab} n_\nu  \, W^\nu_d  W^\mu_c  \nn \\
= \beta \sqrt{g} \left[  ( {\bf n} \cdot {\bf W}_c )  ({\bf X} ^c \cdot \nabla^2 {\bf W} ) 
- K ({\bf n} \cdot {\bf W}_c ) ({\bf n} \cdot {\bf W}^c ) \right]\,.
\ea

The resulting expression for the second variation of the area difference functional (\ref{eq:fmean}) is
\ba
\delta^2 F_{(M)}  [ W ]  =  \int_\mathbb{S}  \sqrt{g} \beta  \{  4 ({\bf X}^a \cdot {\bf  W}^b ) ({\bf n} \cdot {\bf W}_{ab} ) -  2 ({\bf X}^a \cdot {\bf  W}_a ) ( {\bf n} \cdot \nabla^2 {\bf W} )
\nn \\ + 2 ( {\bf n} \cdot {\bf W}_c )  ({\bf X} ^c \cdot \nabla^2 {\bf W} ) 
 - 2 K^{ac} ( {\bf n} \cdot {\bf W}_a ) ({\bf n} \cdot {\bf W}_c )  \nn\\
 +  ( K \, X_{\mu\nu}^{ac} + 2  K_d{}^a X^{cd}_{\mu\nu} - 2 K_d{}^c X^{ad}_{\mu\nu}  ) W^\mu_a W^\nu_c \nn\\
 + 2  K^{cd}  ({\bf X}_c \cdot {\bf W}^a ) ( {\bf X}_d \cdot {\bf W}_a ) 
\}\,.
\ea
This expression reproduces Eq. (101) of \cite{CG04}, obtained via a geometric expansion.
We refrain from presenting the second variation of the Gaussian bending energy, since it is
a total divergence.

The Lagrangian approach described here can be generalized formally in various directions. 
In the presence of external forces, or in the presence of constraints, 
 the variational differential is changed from (\ref{eq:fo1}) to (\ref{eq:fog}),
to include the explicit dependence on the shape functions. The  covariant second variational
derivative (\ref{eq:sd1})  of the energy density is changed accordingly to
\be
\delta^2 \mf = \delta \left[{\partial  \mf \over \partial {\bf X}} \cdot {\bf W} + {\partial \mf \over \partial {\bf X}_a } \cdot {\bf W}_a + {\partial  \mf \over \partial {\bf X}_{ab } } \cdot {\bf W}_{ab}  \right]\,,
\label{eq:fog2}
\ee
where the variational differential of the energy density is given by (\ref{eq:fog}). Clearly,  the
general expression (\ref{eq:svar1}) for the second variation is augmented by the appropriate
Hessians, but with the same general structure.

\section{Concluding remarks}

In this paper, we have developed a fully covariant variational approach to the equilibrium mechanics of fluid lipid membranes, casting it as a covariant Lagrangian classical field theory formulation. As a result, the Euler-Lagrange equations
that define equilibrium take a simple form, in terms of phase space partial derivatives  of the energy density. 
In the same way,  the linear and angular momentum stress tensor are obtained as simple expressions, that 
make clear their interpretation within a classical field theory.
Moreover, the advantages of a Lagrangian formulation become evident when considering issues
of stability, since the second variation of the free energy can be accessed quite readily, and expressed in a form suitable 
for generic applications. 
To bring within the confines of classical field 
theory the study of the equilibrium of arbitrary configurations of lipid membranes, using a language
well established in a physicist's background, may be of help to 
the non-specialist, usually faced with unfamiliar concepts,  and quite often unfamiliar notation as well.

The ambient space covariance is convenient for the formal setting, but more importantly it is needed when 
considering the interaction of the membrane with external forces, and  in particular  when considering hydrodynamic
interactions and viscosity effects, even if remaining in a mean field approximation, where thermal fluctuations are neglected. 
In the study of the dynamics of evolving lipid membranes in the presence of bulk fluids, the
necessity of a covariant approach has  long been recognized and utilized in the derivation 
of full dynamic equations \cite{CL2,hu,Arroyo,powers}. A covariant Lagrangian formulation provides an
alternative  field theoretical point of view .

Finally, we note that various extensions of this work are quite natural.
For example, for the inclusion of the effect of  dissipative fields, a proper Lagrangian formulation allows to introduce, in
appropriate circumstances, a Lagrange-Raleigh variational principle.
It is also  possible to use the Lagrangian formulation to construct a covariant Hamiltonian formulation,
to complement the canonical Hamiltonian formulation of \cite{Ham1}.
Once an Hamiltonian function is available, it is natural also to construct an Hamilton-Jacobi 
formulation, with a  connection of  the theory of  soft surfaces to the  theory of moving 
intefaces \cite{Osher}. Work in these directions is in progress.

\section*{Acknowledgement}

Thanks to Jos\'{e}  M\'{e}ndez and Mart\'{\i}n Hern\'{a}ndez for many useful and motivating conversations, and
to Efra\'in Rojas for his careful reading of this manuscript.  
 Many of the ideas developed in this paper originate from a  collaboration with
 Jemal Guven, and his contribution is gratefully acknowledged.

\nocite{*}

\end{document}